\documentclass{article}
\usepackage[utf8]{inputenc}
\usepackage{amsmath}
\usepackage{mathtools}
\usepackage{amsfonts}
\usepackage{physics}
\usepackage{float}
\usepackage{graphicx}
\usepackage{verbatim}

\usepackage{bbm}
\usepackage{bm}
\usepackage{algorithmic}
\usepackage{algorithm}
\usepackage[round]{natbib}
\usepackage{amsthm} 
\usepackage{algorithmic}
\usepackage{algorithm}
\usepackage{geometry}
\usepackage{mleftright }

\DeclareMathOperator{\cor}{cor}
\DeclareMathOperator{\Var}{Var}

\DeclareMathOperator{\cov}{cov}

\DeclareMathOperator{\E}{E}

\begin{document}



\title{\bfseries \sffamily Bayesian inference for a single factor copula stochastic volatility model using Hamiltonian Monte Carlo}
	\date{\small \today}
			\author{Alexander Kreuzer \footnote{Corresponding author: {E-mail: a.kreuzer@tum.de}}  \and Claudia Czado}
\date{%
	Zentrum Mathematik, Technische Universit\"at M\"unchen\\[2ex]%
	\today
}
			\maketitle
\vspace*{-0.2cm}

\begin{abstract}
For modeling multivariate financial time series we propose a single factor copula model together with stochastic volatility margins. This model generalizes single factor models relying on the multivariate normal distribution and allows for symmetric and asymmetric tail dependence.
We develop joint Bayesian inference using Hamiltonian Monte Carlo (HMC) within Gibbs sampling. Thus we avoid information loss caused by the two-step approach for margins and dependence in copula models as followed by \cite{schamberger2017bayesian}.
Further, the Bayesian approach allows for high dimensional parameter spaces as they are present here in addition to uncertainty quantification through credible intervals.
By allowing for indicators for different copula families the copula families are selected automatically in the Bayesian framework.   
In a first simulation study the performance of HMC is compared to the Markov Chain Monte Carlo (MCMC) approach developed by \cite{schamberger2017bayesian} for the copula part. It is shown that HMC considerably outperforms this approach in terms of effective sample size, MSE and observed coverage probabilities. In a second simulation study satisfactory performance is seen for the full HMC within Gibbs procedure. The approach is illustrated for a portfolio of financial assets with respect to one-day ahead value at risk forecasts. We provide comparison to a two-step estimation procedure of the proposed model and to relevant benchmark models: a model with dynamic linear models for the margins and a single factor copula for the dependence proposed by \cite{schamberger2017bayesian} and a multivariate factor stochastic volatility model proposed by \cite{kastner2017efficient}. Our proposed approach shows superior performance.
\end{abstract}

Keywords: factor copula, stochastic volatility model, Hamiltonian Monte Carlo, value at risk

\section{Introduction}

Multivariate time series models are employed to model the joint behaviour of stocks. It is important to understand the dependence among these financial assets since it has high influence on the performance and the risk associated with a corresponding portfolio (\cite{embrechts2002correlation}, \cite{donnelly2010devil}). Vine copulas (\cite{bedford2001probability}, \cite{aas2009pair}) have proven a useful tool to facilitate complex dependence structures (\cite{nikoloulopoulos2012vine}, \cite{brechmann2013risk}, \cite{aas2016pair}, \cite{fink2017regime}, \cite{nagler2019model}). A vine copula model consists of $\frac{d(d-1)}{2}$pair copulas, where $d$ is the number of assets. So the number of parameters grows quadratically with $d$. \cite{krupskii2013factor} proposed the factor copula model, where the number of parameters grows only linearly in $d$. This model can be seen as a generalization of the Gaussian factor model. The factor copula model provides much more flexibility, compared to the Gaussian one, as it is made up of different pair copulas that can be chosen arbitrarily.
Thus it covers a broad range of dependence structures that can  accommodate symmetric as well as asymmetric tail dependence.

One way to construct multivariate time series models is to combine a univariate time series model for the margins with a dependence model such as the factor copula. Univariate time series models for financial data need to account for typical characteristics like time varying volatility and volatility clustering.
Popular examples of such models include generalized autoregressive conditional heteroskedasticity (GARCH) models (\cite{engle1982autoregressive},  \cite{bollerslev1986generalized}),   the more recently developed generalized autoregressive score (GAS) models (\cite{creal2013generalized}) and stochastic volatility (SV) models (\cite{kim1998stochastic}). Using the classification of \cite{cox1981statistical} GARCH and GAS  models are observation driven models, whereas the SV model is a parameter driven model. In observation driven models volatility is modeled deterministically through the observed past and as such results cannot be transferred to other data sets following the same data generating process. Inference for these observation driven models is often easier since evaluation of the likelihood is straightforward. Inference for SV models is more involved since likelihood evaluation requires high dimensional integration. But efficient MCMC algorithms have been developed (\cite{kastner2014ancillarity}). In the SV model volatility is modeled as latent variables that follow an autoregressive process of order 1. 
This representation has compared favorably to GARCH specifications in several data sets (\cite{yu2002forecasting}, \cite{chan2016modeling}).

We propose a copula based SV model. The marginals follow a SV model and the dependence is modeled through a single factor copula. In contrast to other factor SV models as proposed by  \cite{han2003asset} or \cite{kastner2017efficient} we only allow for one factor and dependence parameters remain constant. But we do not assume that conditional on the  volatilities the observed data is multivariate normal or Student t distributed. Here we provide more flexibility through the choice of different pair copula families.

The single factor copula model has also been deployed by \cite{schamberger2017bayesian} who use dynamic linear models (\cite{west2006bayesian}) as marginals and by \cite{krupskii2013factor} who use GARCH models as marginals.
As it is common in copula modeling, \cite{schamberger2017bayesian} and \cite{krupskii2013factor} both use a two-step approach for estimation. They first estimate marginal parameters and based on these estimates they infer the dependence parameters. \cite{tan2018bayesian} provide full Bayesian inference for a single factor copula based model, but their marginal models have only few parameters and the proposals for MCMC are built using independence among components. However, for SV margins we need to estimate all $T$ log volatilities, where $T$ denotes the length of the time series. Thus we have more than $T$ parameters to estimate per margin. These more sophisticated marginal models for financial data are difficult to handle within a full Bayesian approach. We are able to overcome the two-step approach commonly used in copula modeling  and provide full Bayesian inference. For this we develop and implement a Hamiltonian Monte Carlo (HMC) (\cite{duane1987hybrid}, \cite{neal2011mcmc}) within Gibbs sampler. In HMC information of the gradient of the log posterior density is used to propose new states which leads to an efficient sampling procedure.

 
The main novel contributions of this paper are: joint Bayesian inference of a single factor copula model with SV margins using HMC, automated selection of linking copula families and improved value at risk (VaR) forecasting over benchmark models in a financial application.
More precisely, we first demonstrate how HMC can be employed for the single factor copula model and compare the HMC approach for the copula part to the MCMC approach of \cite{schamberger2017bayesian} who use adaptive rejection Metropolis within Gibbs sampling (\cite{gilks1995adaptive}). HMC shows superior performance in terms of effective sample size, MSE and observed coverage probabilities.  Further, the HMC scheme is integrated within a Gibbs approach that allows for full Bayesian inference of the proposed single factor copula based SV model, including copula family selection. Copula families are modeled with discrete indicator variables, which can be sampled directly from their full conditionals within our Gibbs approach. Continuous parameters are updated with HMC.  Within the Bayesian procedure, marginal and dependence parameters are estimated jointly. We stress that the joint estimation of marginal and dependence parameters is very demanding and is therefore most commonly avoided. Instead a two-step approach is used where the marginal parameters are considered fixed when estimating dependence parameters, i.e. uncertainty in the marginal parameters is ignored. An advantage of the full Bayesian approach is that this uncertainty is not ignored and full uncertainty quantification is straightforward through credible intervals.  We further demonstrate the usefulness of the proposed single factor copula SV model with one-day ahead VaR prediction for financial data involving six stocks. 
Within our full Bayesian approach a VaR forecast is obtained as an empirical quantile of simulations from the predictive distribution. In addition we show that joint estimation leads to more accurate VaR forecasts than VaR forecasts obtained from a two-step approach.

The paper is organized as follows. In Sections \ref{sec:onefact} and \ref{sec:jmod} we discuss the single factor copula model and the single factor copula SV model, respectively. Both sections follow a similar structure. We first specify a Bayesian model, propose a Bayesian inference approach and evaluate the performance of the approach with simulated data. In Section \ref{sec:app} the proposed single factor copula SV model is applied to financial returns data. Section \ref{sec:conc} concludes.

\section{Bayesian inference for single factor copulas using the HMC approach}
\label{sec:onefact}

Hamiltonian dynamics describe the time evolution of a physical system through differential equations. In Hamiltonian Monte Carlo (HMC) the posterior density is connected to the energy function of a physical system. This makes it possible to propose states in the sampling process which are guided by appropriate differential equations. New states are chosen utilizing information of the gradient of the log posterior density, which can lead to more efficient sampling procedures. Therefore HMC has become popular. For example \cite{hartmann2017bayesian} demonstrate how to estimate parameters of generalized extreme value distributions with HMC, while \cite{pakman2014exact} use HMC to sample from truncated multivariate Gaussian distributions. Especially with the development of the probabilistic programming language STAN by \cite{carpenter2016stan} its popularity is growing. STAN allows easy model specification and deploys the No-U-Turn sampler of \cite{hoffman2014no}. This extension of HMC automatically and adaptively selects the tuning parameters. Instead of using STAN we provide our own implementation of HMC. This allows us to use the HMC updates developed for single factor copulas within other samplers, as we will see in Section \ref{sec:jmod}. 
A short introduction to HMC based on \cite{neal2011mcmc} is provided in Appendix 6.1. 
\subsection{Model specification}
To illustrate the viability of HMC for factor copula models we start with the single factor copula model as a special case of the $p$ factor copula model according to \cite{krupskii2013factor}.
We consider $d$ uniform(0,1) distributed variables $U_1, \ldots U_d$ together with a uniform(0,1) distributed latent factor $V$. In the \emph{single factor copula model} we assume that given $V$, the variables $U_1, \ldots, U_d$ are independent. This implies that the joint density of $\boldsymbol U_{1:d} = (U_1, \ldots U_d)^\top$ can be written as
\begin{equation}
\begin{split}
c_{\boldsymbol {U_{1:d}}}(\boldsymbol {u_{1:d}}) &=  \int_0^1 \prod_{j=1}^d c_{j|V}
(u_j|v) dv =\int_0^1 \prod_{j=1}^d c_{j}
(u_j,v) dv, 
\end{split}
\label{onefactordist}
\end{equation}
where $c_{j}$ is the density of $C_j$, the copula of $(U_j,V)$.
The copulas $C_{1}, \ldots,  C_{d}$ are called \emph{linking copulas} as they link each of the observed copula variables $U_j$ to the latent factor $V$.

For inference we use one parametric copula families, i.e. we equip each linking copula density with a corresponding parameter $\theta_j$, and \eqref{onefactordist} becomes

\begin{equation*}
c_{\boldsymbol{U_{1:d}}}(\boldsymbol{u_{1:d}}; \boldsymbol{\theta_{1:d}} ) =  \int_0^1 \prod_{j=1}^d c_{j}
(u_j,v;\theta_j) dv.
\end{equation*}

As it is common in Bayesian statistics we treat the latent variable $V$ as a parameter $v$. The joint density of $\boldsymbol{U_{1:d}}$, denoted by $c_{\boldsymbol{U_{1:d}}}$, given the parameters $(\boldsymbol{\theta_{1:d}} ,v)$ is obtained as

\begin{equation*}
c_{\boldsymbol{U_{1:d}}}(\boldsymbol{u_{1:d}}; \boldsymbol{\theta_{1:d}} ,v) =   \prod_{j=1}^d c_{j}
(u_j,v;\theta_j) .
\end{equation*}

Since the latent variable $V$ is random for each observation vector $(u_{t1}, \ldots, u_{td})^\top$, we have $T$ latent parameters $\boldsymbol v_{1:T} = (v_1, \ldots, v_T)^\top$ for $T$ time points.
The likelihood of the parameters $(\boldsymbol{\theta_{1:d}}, \boldsymbol{v_{1:T}})$ given $T$ independent observations $U_{1:T,1:d}=(u_{tj})_{t = 1, \ldots, T, j=1, \ldots, d}$ is therefore
\begin{equation}
\ell(\boldsymbol{\theta_{1:d}}, \boldsymbol{v_{1:T}}|U_{1:T,1:d}) =   \prod_{t=1}^T\prod_{j=1}^d c_{j}(u_{tj},v_t;\theta_j).
\label{eq:onefaclik}
\end{equation}

\subsection{Bayesian inference}

So far, Bayesian inference for the single factor copula model was addressed by \cite{schamberger2017bayesian} and \cite{tan2018bayesian}. Both approaches use Gibbs sampling where one can exploit the fact that the factors $v_1, \ldots v_T$ are independent given the copula parameters $\theta_1, \ldots \theta_d$ and vice versa. We now show how HMC can be used for the single factor copula model. Sampling with HMC is slower since it requires several evaluations of the gradient of the log posterior density. However with HMC there is no blocking involved and we update the whole parameter vector, with well chosen proposals obtained from the Leapfrog approximation, at once. We expect more accurate samples since this sampler suffers less from the dependence between factors and copula parameters. To support this statement we compare HMC to adaptive rejection Metropolis sampling within Gibbs sampling (ARMGS) (\cite{gilks1995adaptive}). ARMGS is the sampler that worked best among several samplers that have been investigated by \cite{schamberger2017bayesian} for single factor copula models. 

\subsubsection*{Parameterization}

Since HMC operates on unconstrained parameters we need to provide parameter transformations to remove the constraints present in our problem. For many one parametric copula families there is a one-to-one correspondence between the copula parameter $\theta_j$ and Kendall's $\tau$, i.e. there is an invertible function $g_j$ such that $\tau_j = g_j(\theta_j)$, see \cite{joe2014dependence}, Chapter 4. For example  $g_j(\theta_j) =  \frac{2}{\pi} \arcsin\left(\theta_j\right)$ is the corresponding Kendall's $\tau$ for a Gaussian linking copula. Furthermore we restrict the Kendall's $\tau$ values to be in $(0,1)$ to avoid problems that might occur due to multimodal posterior distributions. This is not a severe restriction for applications since $\tau_{U_1,U_2} = -\tau_{U_1,1-U_2}$\footnote{$\tau_{U_1,1-U_2} = P((U_1 - \tilde U_1) ((1-U_2) - (1-\tilde U_2))>0) - P((U_1 - \tilde U_1) ((1-U_2) - (1-\tilde U_2))<0) =  P((U_1 - \tilde U_1) (U_2 -\tilde U_2)<0)            
-P((U_1 - \tilde U_1) (U_2 -\tilde U_2)>0) = - \tau_{U_1,U_2}            
  $, where $(\tilde U_1, \tilde U_2)$ is an independent copy of $(U_1,U_2)$.}. So we can replace $U_2$ by $1-U_2$ if we want to model negative dependence between $U_1$ and $U_2$. The components of the latent factors $\boldsymbol{v_{1:T}}$ are also in $(0,1)$. To transform parameters on the $(0,1)$ scale to the unconstrained scale the logit function is a common choice. Therefore we use the following transformations for the copula parameters $\boldsymbol {\theta_{1:d}}$ and the latent factors $\boldsymbol{v_{1:T}}$
\begin{equation}
\delta_j = \ln(\frac{g_j(\theta_j)}{1-g_j(\theta_j)}),~~~
w_t = \ln(\frac{v_t}{1-v_t}),
\end{equation}
and obtain unconstrained parameters $\delta_j,v_t \in \mathbb{R}$ for $ j = 1,\ldots d, t=1, \ldots,T$. 

\subsubsection*{Prior densities}

We specify the prior distributions for $(\boldsymbol{\delta_{1:d}}, \boldsymbol{w_{1:T}})$ such that the distributions implied for the corresponding Kendall's $\tau$ and for $v_t$ are independently uniform on the interval $(0,1)$. Applying the density transformation law this implies that the factor copula ($FC$) prior density can be expressed as
\begin{equation} 
\begin{split}
\pi_{FC}(\boldsymbol{\delta_{1:d}},  \boldsymbol{w_{1:T}}) &= \prod_{t=1}^T \pi_u(w_t) \prod_{j=1}^d \pi_u(\delta_j), \\
\end{split}
\label{eq:priorfc}
\end{equation}
where $\pi_{u}(x) = (1+\exp(-x))^{-2} \exp(-x), x \in \mathbb{R}.$
\subsubsection*{Posterior density}

With these choices in \eqref{eq:onefaclik} and \eqref{eq:priorfc} the posterior density is proportional to
\begin{equation}
\begin{split}
f(\boldsymbol{\delta_{1:d}}, \boldsymbol{w_{1:T}}|U_{1:T,1:d}) \propto 
 l(\boldsymbol{\theta_{1:d}}, \boldsymbol{v_{1:T}}|U_{1:T,1:d})  \cdot \pi_{FC}(\boldsymbol{\delta_{1:d}},\boldsymbol{w_{1:T}}),\\
\end{split}
\label{eq:onefacpost}
\end{equation}
where $\theta_j$ and $v_t$ are functions of $\delta_j$ and $w_t$ respectively. Therefore the log posterior density is, up to an additive constant, given by

\begin{equation*}
\begin{split}
\mathcal{L}(\boldsymbol{\delta_{1:d}},\boldsymbol{w_{1:T}}|U_{1:T,1:d}) \propto&  \sum_{t=1}^T \sum_{j=1}^d \ln(c_j(u_{tj},v_t;\theta_j))+    \sum_{t=1}^T \ln(\pi_u(w_t))+
\sum_{j=1}^d \ln(\pi_u(\delta_j)).
\end{split}
\end{equation*}

\subsubsection*{Sampling with HMC}
Derivatives of the log posterior density with respect to all parameters are determined to perform Leapfrog approximations (see Appendix \ref{seq:deriv_hmc}). With this at hand, HMC can be implemented as any Metropolis-Hastings sampler. To run the algorithm we need to set values to the hyper parameters: the Leapfrog stepsize $\epsilon$, the number of Leapfrog steps $L$ and the mass matrix $M$. Choosing $\epsilon$ and $L$ is not easy since good choices of  these parameters can vary depending on different regions of the state space.  \cite{neal2011mcmc} suggest to randomly select $\epsilon$ and $L$ from a set of values that may be appropriate for different regions. This is the approach that we follow. For our simulation study we have seen that choosing $\epsilon$ uniformly between $0$ and $0.2$ and choosing $L$ uniformly between $0$ and $40$ leads to reasonable mixing as measured by the effective sample size (\cite{gelman2014bayesian}, page 286). The mass matrix $M$ is set equal to the identity matrix.
The MCMC procedure is implemented in $\texttt{R}$ using the $\texttt{R}$ package $\texttt{Rcpp}$ by \cite{eddelbuettel2011rcpp} which allows the integration of $\texttt{C++}$. Effective sample sizes are calculated with the $\texttt{R}$ package $\texttt{coda}$ by \cite{plummer2008coda}.

\subsection{Simulation study}
\label{sec:simonefac}
To compare our approach we conduct the same simulation study as in \cite{schamberger2017bayesian}. For each of three scenarios, we simulate 100 data sets from the single factor copula model with $T=200$ and $d=5$. The three scenarios are characterized by the values of Kendall's $\tau$ of the linking copulas and are denoted by  the \textbf{low $\tau$}, the \textbf{high $\tau$} and the \textbf{mixed $\tau$} scenario. The Kendall's $\tau$ values are shown in Table \ref{tausimstudy}. As linking copulas only Gumbel copulas are considered. 
Based on  these simulated data sets the samplers are run for 11000 iterations, whereas the first 1000 iterations are discarded for burn in.  

\begin{table}[ht]
\scriptsize
\centering
\begin{tabular}{rrrrrr}
  \hline
 & $C_{1}$ & $C_{2}$ & $C_{3}$ & $C_{4}$ & $C_{5}$ \\ 
  \hline
\textbf{low $\tau$}& 0.10 & 0.12 & 0.15 & 0.18 & 0.20 \\ 
\textbf{high $\tau$} & 0.50 & 0.57 & 0.65 & 0.73 & 0.80 \\ 
\textbf{mixed $\tau$} & 0.10 & 0.28 & 0.45 & 0.62 & 0.80 \\ 
   \hline
\end{tabular}
\caption{Kendall's $\tau$ values for the linking copulas $C_{1}, \ldots C_{5}$ in the three scenarios.}
\label{tausimstudy}
\end{table}

Table \ref{schsimstudy} shows the results of the simulation study and compares them to the results obtained by \cite{schamberger2017bayesian} using adaptive rejection Metropolis sampling within Gibbs sampling (ARMGS). The corresponding error statistics (e.g. mean absolute deviation (MAD), mean squared error (MSE)) for each parameter is obtained from 100 replications.  Then, e.g. the MSE for $\tau$ in Table \ref{schsimstudy} is computed as the average of MSE for $\tau_1$ $,\ldots,$ MSE for $\tau_5$. Since the objective is the comparison of our method to the method of \cite{schamberger2017bayesian} we follow their approach and calculate the error statistics from point estimates (marginal posterior mode estimates  obtained as the estimated modes of univariate kernel density estimates). Further we calculate the error statistics for $\boldsymbol \tau_{1:d}, \boldsymbol  v_{1:T}$ which are one-to-one transformations of $\boldsymbol \delta_{1:d}, \boldsymbol w_{1:T}$.   
 We see that a more accurate credible interval, a lower mean absolute deviation and a lower mean squared error is achieved in most cases by HMC compared to ARMGS. Furthermore the effective sample size per minute is much higher for HMC. Table \ref{tab:detsimscham} shows the results of the simulation study in more detail, i.e. we do not average over values of $\tau_1, \ldots, \tau_5$ and $v_1, \ldots v_{200}$. It is noticable that mixing is worse for higher values of Kendall's $\tau$ in every scenario, whereas it is most extreme in the mixed $\tau$ scenario. This was also observed for ARMGS (see \cite{schamberger2017bayesian} Table 9 in the appendix). 

\begin{table}[H]
\scriptsize
\centering
\begin{tabular}{lll|ll}
&\multicolumn{2}{c}{$\tau$} & \multicolumn{2}{c}{$V$} \\
  \hline
 & ARMGS & HMC & ARMGS & HMC \\ 
  \hline
\textbf{Low $\tau$} &&&&\\
MAD & 0.1088 & 0.0564 & 0.2808 & 0.2158 \\ 
MSE & 0.0314 & 0.0059 & 0.1248 & 0.0716 \\ 
ESS/min & 6 & 92 & 26 & 246 \\ 
90$\%$ C.I. & 0.91 & 0.94 & 0.84 & 0.88 \\ 
95$\%$ C.I. & 0.96 & 0.98 & 0.91 & 0.94 \\ 
\textbf{High $\tau$}&&&&\\
MAD & 0.0292 & 0.0201 & 0.0709 & 0.0502 \\ 
 MSE & 0.0014 & 0.0007 & 0.0095 & 0.0046 \\ 
ESS/min & 24 & 268 & 44 & 278 \\ 
  90$\%$ C.I. & 0.89 & 0.90 & 0.89 & 0.91 \\ 
95$\%$ C.I.  & 0.95 & 0.94 & 0.95 & 0.95 \\ 
\textbf{Mixed $\tau$}&&&&\\
MAD & 0.0509 & 0.0340 & 0.0828 & 0.0684 \\ 
MSE & 0.0043 & 0.0019 & 0.0132 & 0.0082 \\ 
ESS/min & 21 & 132 & 26 & 210 \\ 
 90$\%$ C.I. & 0.87 & 0.89 & 0.79 & 0.85 \\ 
 95$\%$ C.I. & 0.93 & 0.93 & 0.88 & 0.93 \\ 
   \hline
\end{tabular}
\caption{Comparison of the ARMGS and HMC method in terms of mean absolute deviation (MAD), mean squared error (MSE), effective sample size per minute (ESS/min) and observed coverage probability of the credible intervals (C.I.).}
\label{schsimstudy}
\end{table}

\begin{table}[H]
\scriptsize
\centering
\begin{tabular}{lllllllllll}
  \hline
 & $\tau_1$ & $\tau_2$ & $\tau_3$ & $\tau_4$ & $\tau_5$ & $v_{10}$ & $v_{50}$ & $v_{100}$ & $v_{150}$ & $v_{190}$ \\ 
  \hline
\textbf{Low $\tau$} &&&&&&&&&&\\
MAD & 0.0500 & 0.0493 & 0.0568 & 0.0591 & 0.0666 & 0.2530 & 0.2204 & 0.2376 & 0.2190 & 0.2203 \\ 
MSE & 0.0047 & 0.0043 & 0.0049 & 0.0058 & 0.0096 & 0.0969 & 0.0719 & 0.0819 & 0.0712 & 0.0703 \\ 
ESS/min & 121 & 114 & 91 & 76 & 58 & 239 & 255 & 268 & 247 & 262 \\ 
90$\%$ C.I. & 0.94 & 0.94 & 0.91 & 0.92 & 0.98 & 0.85 & 0.88 & 0.83 & 0.88 & 0.90 \\ 
95$\%$ C.I. & 0.98 & 0.98 & 0.97 & 0.96 & 0.99 & 0.93 & 0.96 & 0.90 & 0.91 & 0.93 \\ 
\textbf{High $\tau$} &&&&&&&&&&\\
MAD & 0.0253 & 0.0214 & 0.0204 & 0.0158 & 0.0174 & 0.0549 & 0.0549 & 0.0475 & 0.0474 & 0.0474 \\ 
MSE & 0.0011 & 0.0007 & 0.0007 & 0.0004 & 0.0005 & 0.0058 & 0.0049 & 0.0043 & 0.0043 & 0.0044 \\ 
ESS/min  & 320 & 319 & 312 & 265 & 125 & 277 & 278 & 275 & 279 & 278 \\ 
90$\%$ C.I. & 0.84 & 0.91 & 0.88 & 0.92 & 0.94 & 0.86 & 0.86 & 0.88 & 0.95 & 0.90 \\ 
95$\%$ C.I.& 0.89 & 0.95 & 0.94 & 0.97 & 0.97 & 0.92 & 0.96 & 0.94 & 0.96 & 0.95 \\ 
 \textbf{Mixed $\tau$} &&&&&&&&&&\\
MAD & 0.0375 & 0.0358 & 0.0278 & 0.0256 & 0.0431 & 0.0690 & 0.0668 & 0.0636 & 0.0712 & 0.0663 \\ 
MSE & 0.0021 & 0.0020 & 0.0012 & 0.0010 & 0.0031 & 0.0098 & 0.0078 & 0.0080 & 0.0084 & 0.0085 \\ 
ESS/min & 147 & 250 & 201 & 50 & 11 & 212 & 207 & 224 & 218 & 221 \\ 
90$\%$ C.I.& 0.89 & 0.85 & 0.88 & 0.93 & 0.89 & 0.87 & 0.87 & 0.87 & 0.81 & 0.87 \\ 
95$\%$ C.I. & 0.95 & 0.89 & 0.94 & 0.94 & 0.94 & 0.89 & 0.94 & 0.94 & 0.88 & 0.92 \\ 
   \hline
\end{tabular}
\caption{Detailed simulation results for the HMC method. We show the estimated mean absolute deviation (MAD), mean squared error (MSE), effective sample size per minute (ESS/min) and observed coverage probability of the credible intervals (C.I.) for $\tau_1, \ldots, \tau_5$ and five selected latent variables $v_t, t = 10,50,100,150,190$.}
\label{tab:detsimscham}
\end{table}

\section{The single factor copula stochastic volatility model}
\label{sec:jmod}
Now we combine the single factor copula with margins driven by a stochastic volatility model and develop a Bayesian approach to jointly estimate the parameters of the proposed model.

\subsection{Model specification}

\label{sec:modelspec}

\subsubsection*{The marginal model}
We utilize the stochastic volatility model (\cite{kim1998stochastic}) as marginal model. In this model the log variances $(s_1, \ldots, s_T)^\top$ of a conditionally normally distributed vector $(Z_1, \ldots, Z_T)^\top$ are modeled with a latent AR(1) process. This AR(1) process has mean parameter $\mu \in \mathbb{R}$, persistence parameter $\phi\in (-1,1)$ and standard deviation parameter $\sigma \in (0,\infty)$. More precisely, the \emph{stochastic volatility (SV) model} is given by

\begin{equation}
\begin{split}
Z_t &= \exp(\frac{s_t}{2}) \epsilon_t, ~~ t = 1, \ldots T, \\
s_t &= \mu + \phi (s_{t-1} - \mu) + \sigma \eta_t , ~~ t = 1, \ldots T,
\end{split}
\end{equation}
where $s_0|\mu, \phi, \sigma \sim N\left(\mu, \frac{\sigma^2}{1- \phi^2}\right)$ and $\epsilon_t, \eta_t, \sim N(0,1)$ independently, for $t =1, \ldots, T$.

\cite{kastner2014ancillarity} develop an MCMC algorithm for this model which uses the ancillarity-sufficiency interweaving strategy proposed by \cite{yu2011center}. This strategy leads to an efficient MCMC sampling procedure which is implemented in the $\texttt{R}$ package  $\texttt{stochvol}$ (see \cite{kastner2016dealing}).
We discuss the prior densities proposed by \cite{kastner2016dealing} since we also utilize them later.
The following priors for $\mu, \phi$ and $\sigma$ are chosen
\begin{equation}
\begin{split}
&\mu \sim N(0,100), ~~ \frac{\phi + 1}{2} \sim Beta(5,1.5), ~~\sigma^2 \sim \chi^2_1.
\end{split}
\label{eq:svprior}
\end{equation}

 The prior for $\mu$ is rather uninformative, whereas the prior for $\phi$ puts more mass on higher values for the persistence parameter. High persistence parameters are characteristic for financial time series. The prior choice for $\sigma^2$ differs from the popular inverse Gamma prior. In contrast to the inverse Gamma prior, the $\chi^2_1$ has more mass close to zero and thus allows for latent volatilities with less fluctuations.   
Denoting by $\boldsymbol s_{0:T} =(s_0, \ldots, s_T)^\top$ the vector of latent log variances, the prior density of $(\mu, \phi, \sigma, \boldsymbol {s_{0:T}}^\top)^\top$ is given by
\begin{equation}
\begin{split}
\pi_{SV}(\mu, \phi, \sigma, \boldsymbol {s_{0:T}}) &= f( \boldsymbol {s_{0:T}}|\mu, \phi, \sigma) f(\mu, \phi, \sigma)\\
&= \varphi\left(s_0 \Big|\mu, \frac{\sigma^2}{1-\phi^2}\right) \prod_{t=1}^T \varphi\left(s_t|\mu + \phi(s_{t-1}-\mu),\sigma^2\right) \pi(\mu) \pi(\phi) \pi(\sigma),
\end{split}
\label{eq:pisv}
\end{equation}
where $\varphi\left(\cdot|\mu_{normal},\sigma_{normal}^2\right)$ denotes a univariate normal density with mean $\mu_{normal}$ and variance $\sigma_{normal}^2$ and $\pi(\cdot)$ denotes the corresponding prior density as specified in \eqref{eq:svprior}.

\subsubsection*{The joint model}

We propose a multivariate dynamic model where each marginal follows a stochastic volatility model and the dependence between the marginals is captured by a single factor copula, the \emph{single factor copula stochastic volatility (factor copula SV) model}. 
In particular for $t=1,\ldots, T, j=1, \ldots,d$ we assume that
\begin{equation*}
\begin{split}
Z_{tj} &= \exp(\frac{s_{tj}}{2}) \epsilon_{tj} \\
s_{tj} &= \mu_j + \phi_j ( s_{t-1j} - \mu_j) + \sigma_j \eta_{tj},
\end{split}
\end{equation*}
where $\mu_j \in \mathbb{R}, \phi_j \in (-1,1), \sigma_j \in (0, \infty),  s_{0j}|\mu_j, \phi_j, \sigma_j \sim N\left(\mu_j, \frac{\sigma_j^2}{1- \phi_j^2}\right)$ and $\eta_{tj} \sim N(0,1)$ i.i.d. holds. 
The joint distribtion of the errors $\epsilon_{tj}$ is now considered. We model the dependence among the marginals by employing a factor copula model on the errors. We further allow for Bayesian selection of the $d$ linking copula families of this factor copula instead of assuming that they were known as in Section \ref{sec:onefact} and as in \cite{schamberger2017bayesian}. The families are chosen from a set $\mathcal{M}$ of one parametric copula families, e.g. $\mathcal{M} = \{\text{Gaussian, Gumbel, Clayton}\}$. \cite{schamberger2017bayesian} estimated one model for each specification of the linking copula families. Since there are $|\mathcal{M}|^d$ different specifications, they only considered factor copulas where all linking copulas belong to the same family.  With our Bayesian family selection, we can profit from the full flexibility of the factor copula model by allowing for all $|\mathcal{M}|^d$ specifications. In particular our modeling approach allows to combine different copula families. Therefore we define $d$ family indicator variables $m_j \in \mathcal{M}, j=1,\ldots, d $.  Further, we introduce parameters $\delta_j \in \mathbb{R}, j=1, \ldots, d$ which are mapped to the corresponding Kendall's $\tau$ with the sigmoid (inverse logit) transform. This Kendall's $\tau$ is then mapped  to the corresponding copula parameter with the function $g^{-1}_{m_j}$, i.e. 
\begin{equation}
\theta_j^{m_j} = \theta_j^{m_j}(\delta_j) = g^{-1}_{m_j}\left(\frac{\exp(\delta_j)}{1+\exp(\delta_j)}\right).
\end{equation}
Note that the parameter $\delta_j$ has the same interpretation for different copula families: It is the logit transform of the associated Kendall's $\tau$ value. This allows to share this parameter among different copula families. In the following, the with copula family $m_j$ associated copula parameter  $\theta_j^{m_j}$ is  determined as a function of $\delta_j$ and $m_j$.
Since we model the dependence among the errors with a single factor copula, we assume that there exists a latent factor $v_t \sim unif(0,1)$ for each $t$ such that the following holds for the error vector at time $t$, $\boldsymbol \epsilon_{t\cdot} = (\epsilon_{t1}, \ldots \epsilon_{td})^\top$,
\begin{equation}
f(\boldsymbol \epsilon_{t\cdot}|v_t, \boldsymbol m_{1:d}, \theta_1^{m_1}, \ldots, \theta_d^{m_d}) = \prod_{j=1}^d \Big[c^{m_j}_j(\Phi(\epsilon_{tj}),v_t;\theta_j^{m_j}) \varphi(\epsilon_{tj})\Big],
\label{eq:epsdef}
\end{equation}
where $\Phi$ denotes the standard normal distribution function. In particular $\epsilon_{tj} \sim N(0,1)$ for any $t$ and $j$.
Here $c^{m_j}_j(\cdot,\cdot;\theta_j^{m_j})$ is the density of the  bivariate copula family $m_j$ with parameter $\theta_j^{m_j}$. Integrating out the factor $v_t$ in \eqref{eq:epsdef} yields
\begin{equation}
f(\boldsymbol \epsilon_{t\cdot}|\boldsymbol m_{1:d}, \theta_1^{m_1}, \ldots, \theta_d^{m_d}) = \Bigg[\int_{(0,1)} \prod_{j=1}^d c^{m_j}_j(\Phi(\epsilon_{tj}),v_t;\theta_j^{m_j}) dv_t \Bigg] \prod_{j=1}^d \varphi(\epsilon_{tj}).
\label{eq:epsuncond}
\end{equation}

Furthermore we assume that the $T$ components of ($\boldsymbol \epsilon_{1\cdot}$, $\ldots$, $\boldsymbol \epsilon_{T\cdot}$) are independent  given the family indicators $\boldsymbol m_{1:d}$ and the dependence parameters $\boldsymbol \delta_{1:d}, \boldsymbol v_{1:T}$. 
To shorten notation we use the following abbreviations: 
\begin{itemize}
\item $Z = (z_{tj})_{t=1,\ldots, T, j=1,\ldots d}$ the matrix of observations,
\item $\mathcal{E} = (\epsilon_{tj})_{t=1,\ldots, T, j=1,\ldots d}$ the matrix of errors,
\item $\boldsymbol \mu = (\mu_j)_{j=1,\ldots d}$ the vector of means of the marginal stochastic volatility models,
\item $\boldsymbol \phi = (\phi_j)_{j=1,\ldots d}$ the vector of persistence parameters of the marginal stochastic volatility models,
\item $\boldsymbol \sigma = (\sigma_j)_{j=1,\ldots d}$ the vector of standard deviations of the marginal stochastic volatility models,
\item $S = (s_{tj})_{t=0,\ldots, T, j=1,\ldots d}$ the matrix of log variances,
\item $\boldsymbol s_{\cdot j} = (s_{tj})_{t=0,\ldots, T}$ the vector of log variances of the $j$-th marginal,
\item $\boldsymbol v = (v_t)_{t=1,\ldots, T}$ the vector of latent factors,
\item $\boldsymbol \delta = (\delta_j)_{j=1,\ldots, d}$ the vector of transformed copula parameters,
\item $\boldsymbol m = (m_j)_{j=1,\ldots, d}$ the vector of copula family indicators.
\end{itemize}
Utilizing these abbreviations, we can summarize the parameters of our model as $\{\boldsymbol \mu, \boldsymbol \phi, \boldsymbol \sigma, S, \boldsymbol v, \boldsymbol \delta, \boldsymbol m  \}.$

\subsubsection*{The model with Gaussian linking copulas}
For the special case where all linking copulas are Gaussian (i.e. $m_j =$ Gaussian for $j = 1, \ldots, d$), \cite{krupskii2013factor} show that the errors $\epsilon_{tj}$ specified in \eqref{eq:epsuncond} allow for the following stochastic representation
\begin{equation*}
\epsilon_{tj} = \rho_j w_t + \sqrt{1-\rho_j^2} \xi_{tj},
\end{equation*}
where $w_t \sim N(0,1)$ and $\xi_{tj} \sim N(0,1)$ independently and $\rho_j = \theta_j^{m_j}$ is the Gaussian copula parameter. Therefore we obtain the following additive structure
\begin{equation}
Z_{tj} =  \rho_j \exp(\frac{s_{tj}}{2}) w_t + \exp(\frac{s_{tj}}{2})  \sqrt{1-\rho_j^2} \xi_{tj}.
\label{eq:addstruct}
\end{equation}

This implies a time dynamic covariance matrix with elements
\begin{equation*}
\cov(Z_{tj},Z_{tk}) =  \rho_j \rho_k \exp(\frac{s_{tj}}{2}) \exp(\frac{s_{tk}}{2}) \text{ for } j \neq k. 
\end{equation*}
The correlation matrix however remains constant as time evolves and its off-diagonal elements are given by
\begin{equation*}
\cor(Z_{tj},Z_{tk}) =  \rho_j \rho_k  \text{ for } j \neq k.
\end{equation*}
The additive structure in \eqref{eq:addstruct} shows connections to other multivariate factor stochastic volatility models (see \cite{chib2006analysis}, \cite{kastner2017efficient}). This can be seen by considering the following reparameterization
\begin{equation*}
\begin{split}
s_{tj}' &\coloneqq {s_{tj}} + \ln(1-\rho_{j}^2), \lambda_j \coloneqq \frac{\rho_j}{\sqrt{1-\rho_j^2}}, \\  
\end{split}
\end{equation*}
which implies the following representation of \eqref{eq:addstruct}
\begin{equation}
Z_{tj} = \lambda_j \exp(\frac{s_{tj}'}{2}) w_t + \exp(\frac{s_{tj}'}{2}) \xi_{tj}.
\label{eq:newrep}
\end{equation}
Here $s_{tj}'$ is an AR(1) process with mean $\mu_j + \ln(1-\rho_j^2)$, persistence parameter $\phi_j$ and standard deviation parameter $\sigma_j$.

For comparison, the model of \cite{kastner2017efficient} with one factor is given by
\begin{equation*}
Z_{tj} = \lambda_j \exp(\frac{s_{td+1}'}{2}) w_t + \exp(\frac{s_{tj}'}{2}) \xi_{tj},
\end{equation*}
with one additional latent AR(1) process $s_{td+1}', t=1, \ldots, T$. This implies time varying correlations given by
\begin{equation*}
\cor(Z_{tj},Z_{tk})=\frac{\lambda_j\lambda_k \exp(s_{td+1}')}{\sqrt{\lambda_j^2\exp(s_{td+1}')+\exp(s_{tj}')}\sqrt{\lambda_k^2\exp(s_{td+1}')+\exp(s_{tk}')}} \text{ for } j \neq k.
\end{equation*}

Dividing $Z_{tj}$ by $\exp(\frac{s_{tj}'}{2})$ in \eqref{eq:newrep} we recognize the structure of a standard factor model for $Z_{tj}' \coloneqq \frac{Z_{tj}}{\exp(\frac{s_{tj}'}{2})}$ given by
\begin{equation}
Z_{tj}' = \lambda_j  w_t + \xi_{tj},
\label{eq:stand_fac}
\end{equation}
with factor loadings $\lambda_1, \ldots, \lambda_d$ and factor $w_t$. In representation \eqref{eq:stand_fac} the variance of $\xi_{tj}$ is restricted to 1 whereas in the standard factor model (see e.g. \cite{lopes2004bayesian}) it is usually modeled through an additional variance parameter. Since the variance of $\epsilon_{tj}$ is already determined ($\epsilon_{tj} \sim N(0,1)$) we have this additional restriction compared to factor models with flexible marginal variance. Note that $Z_{tj}$ still has flexible variance and the restriction for $\epsilon_{tj}$ is necessary to ensure identifiability.

If all copula families are Gaussian other multivariate factor stochastic volatility  models provide generalizations by allowing for more factors and for a time varying correlation. We provide generalization with respect to the error distribution. The choice of  different pair copula families provides a flexible modeling approach and our model can accommodate features that can not be modeled with a multivariate normal distribution as e.g. symmetric or asymmetric tail dependence.

\cite{schamberger2017bayesian} also use factor copulas to model dependence among financial assets. Their approach differs to our approach in the choice of the marginal model. They use dynamic linear models (\cite{west2006bayesian}). Secondly they assume the copula families to be known and they perform a two-step estimation approach, whereas we provide full Bayesian inference.

\subsection{Bayesian inference}
In the following we develop a full Bayesian approach for the proposed model.  We use a block Gibbs sampler to sample from the posterior distribution. We use $d$ blocks for the marginal parameters  $(\mu_j, \phi_j , \sigma_j, \boldsymbol s_{\cdot j})$, $j=1, \ldots,d$, one block for the dependence parameters $(\boldsymbol \delta, \boldsymbol v)$ and $d$ blocks for the copula family indicators $\boldsymbol m$. Sampling from the full conditionals is done with HMC for the first $d+1$ blocks. Conditioning the dependence parameters on the marginal parameters and on the copula family indicators we are in the single factor copula framework of Section \ref{sec:onefact}. We have seen that HMC provides an efficient way to sample the dependence parameters. Conditioned on the dependence parameters  and on the family indicators, the marginal parameters corresponding to different dimensions are independent. Each dimension can be considered as a generalized stochastic volatility model, where the distribution of the errors is determined by the corresponding linking copula. Sampling from the posterior distribution is more involved than in the Gaussian case. In the Gaussian case one can use an approximation of a mixture of normal distributions and rewrite the observation equation $Z_{tj} = \exp(\frac{s_{tj}}{2}) \epsilon_{tj}$  as a linear, conditionally Gaussian state space model (\cite{omori2007stochastic}, \cite{kastner2014ancillarity}). This is not possible in our case and therefore HMC, which has already shown good performance for the copula part and only requires derivation of the derivatives, is our method of choice.
The family indicators $\boldsymbol m$ are discrete variables which can be sampled directly from their full conditionals. 

\label{sec:bimod}
\subsubsection*{Prior densities}
For the copula family indicators we use independent discrete uniform priors, i.e
\begin{equation}
\pi(m_j) = \frac{1}{|\mathcal{M}|}
\end{equation}
for  $m_j \in \mathcal{M}, j=1, \ldots, d$ independently.
The prior density of the other parameters is chosen as the product of the priors used for the marginal stochastic volatility model and for the single factor copula model, i.e.
\begin{equation}
\pi_J(\boldsymbol\mu, \boldsymbol\phi, \boldsymbol\sigma, S, \boldsymbol\delta,\boldsymbol v) = \prod_{j=1}^d \pi_{SV}(\mu_j, \phi_j,\sigma_j, \boldsymbol s_{\cdot j}) \pi_{u}(\delta_j),
\label{eq:priorfull}
\end{equation}
where $\pi_{SV}(\cdot)$ and $\pi_{u}(\cdot)$ are specified in \eqref{eq:pisv} and \eqref{eq:priorfc}, respectively. Further we assume that the family indicators are a priori independent of the parameters in \eqref{eq:priorfull}.

\subsubsection*{Likelihood}
The conditional independence of the $T$ components of ($\boldsymbol \epsilon_{1\cdot}$, $\ldots$, $\boldsymbol \epsilon_{T\cdot}$) implies that the conditional distribution of the errors given the dependence parameters and the copula family indicators is 
 \begin{equation*}
 f(\mathcal{E}|\boldsymbol v, \boldsymbol \delta, \boldsymbol m) = \prod_{t=1}^T \prod_{j=1}^d \Big[c^{m_j}_j(\Phi(\epsilon_{tj}),v_t;\theta_j^{m_j}) \varphi(\epsilon_{tj})\Big].
 \end{equation*}

Using the density transformation rule, the likelihood of parameters  $(\boldsymbol\mu, \boldsymbol\phi, \boldsymbol\sigma, S, \boldsymbol\delta,\boldsymbol v, \boldsymbol m)$ given the observation matrix $Z$  is obtained as

\begin{equation*}
\begin{split}
\ell(\boldsymbol\mu, \boldsymbol\phi, \boldsymbol\sigma, S, \boldsymbol\delta,\boldsymbol v, \boldsymbol m|Z) &= \prod_{t=1}^T\prod_{j=1}^d \left[ c_j^{m_j}\mleft(\Phi\left(\frac{z_{tj}}{\exp(\frac{s_{tj}}{2})}\right),v_t;\theta_j^{m_j} \mright) \varphi\mleft(\frac{z_{tj}}{\exp(\frac{s_{tj}}{2})}\mright)  \frac{1}{\exp(\frac{s_{tj}}{2})} \right] .\\
\end{split}
\end{equation*}

\subsubsection*{Sampling the marginal parameters}
The conditional density we need to sample from is given by
\begin{equation*}
\begin{split}
f(\mu_j, \phi_j, &\sigma_j, \boldsymbol s_{\cdot j}|  Z,\boldsymbol\mu_{-j},\boldsymbol\phi_{-j},\boldsymbol\sigma_{-j},S_{\cdot -j},\boldsymbol\delta,\boldsymbol v, \boldsymbol m)\\
 &\propto \ell(\boldsymbol\mu, \boldsymbol\phi, \boldsymbol\sigma, S, \boldsymbol\delta,\boldsymbol v, \boldsymbol m|Z) \pi_J(\boldsymbol\mu, \boldsymbol\phi, \boldsymbol\sigma, S, \boldsymbol\delta, \boldsymbol v)\\
 &\propto  \prod_{t=1}^T \left[ c_j^{m_j}\mleft(\Phi\mleft(\frac{z_{tj}}{\exp(\frac{s_{tj}}{2})}\mright),v_t;\theta_j^{m_j} \mright) \varphi\mleft(\frac{z_{tj}}{\exp(\frac{s_{tj}}{2})}\mright)  \frac{1}{\exp(\frac{s_{tj}}{2})} \right]  \pi_{SV}(\mu_j, \phi_j,\sigma_j, \boldsymbol s_{\cdot j}). \\
\end{split}
\end{equation*}

Here the abbreviation $\boldsymbol x_{-j}$ refers to the vector $(x_1, \ldots x_d)^\top$ with the $j-$th component removed and $X_{\cdot -j}$ is the matrix $X$ with the $j$-th column removed.
We sample from this density with HMC as will be outlined below.\vspace*{0.2cm}

\textit{Parameterization} \hspace*{0.2cm} As in Section \ref{sec:onefact} we need to provide parameterizations such that resulting parameters are unconstrained. In particular we use the following transformations
\begin{equation*}
\begin{split}
\xi_j &= F_Z(\phi_j), ~~~~~ \psi_j =\ln(\sigma_j), \\
\end{split}
\label{eq:partrafo1}
\end{equation*}
where $F_Z(x) = \frac{1}{2}\ln(\frac{1+x}{1-x})$ is Fisher's Z transformation. Although the latent log variances are already unconstrained  we make use of the following reparameterization 
\begin{equation}
\begin{split}
\tilde s_{0j} &= \frac{(s_{0j}-\mu_j)\sqrt{1-\phi_j^2}}{\sigma_j} \\
\tilde s_{tj} &= \frac{s_{tj}-\mu_j-\phi_j(s_{t-1j}-\mu_j)}{\sigma_j}, t=1, \ldots, T. \\
\end{split}
\label{eq:partrafo}
\end{equation}
The transformation for $\boldsymbol {s_{\cdot j}}$ was proposed by the Stan \cite{team2015stan} for the univariate stochastic volatility model and implies that $\boldsymbol {\tilde s_{\cdot j}}|\mu_j,\phi_j,\sigma_j \sim N(0, I_{T+1})$, where $I_{T+1}$ denotes the $(T+1)$-dimensional identity matrix. According to \cite{yu2011center} the original parameterization in terms of $s_{tj}$ is a sufficient augmentation scheme, whereas the parameterization in terms of $\tilde s_{tj}$ is  an ancillary augmentation. The performance of Markov Chain Monte Carlo methods can vary a lot for different parameterizations (\cite{fruhwirth2003bayesian}, \cite{strickland2008parameterisation}).
\cite{betancourt2015hamiltonian} have seen better performance for the ancillary augmentation when sampling from the posterior distribution of hierarchical models with HMC. Their explanation is that within the ancillary augmentation variables may be less correlated. Here we also rely on the ancillary augmentation since we  have seen much better performance for this parameterization in terms of effective sample size.

\vspace*{0.2cm}
\textit{Prior densities}  \hspace*{0.2cm} 
We denote by $\pi_{SV2}$ the joint prior density of the parameters $\mu_j, \xi_j, \psi_j$ and $\boldsymbol {\tilde s_{\cdot j}}$.
The log of this joint prior density is, up to an additive constant, given by
\begin{equation*}
\ln(\pi_{SV2}(\mu_j, \xi_j, \psi_j,\boldsymbol{\tilde s}_{\cdot j})) \propto \ln(\pi(\mu_j)) + \ln(\pi(\xi_j)) + \ln(\pi(\psi_j)) - \frac{1}{2} \sum_{t=0}^{T} \tilde s_{tj}^2.
\end{equation*}
where $\pi(\cdot)$ are the corresponding prior densities implied by \eqref{eq:priorfull} (see Appendix \ref{sec:trafo_dens} for details).

\vspace*{0.2cm}
\textit{Posterior density} \hspace*{0.2cm}  The log posterior density we need to sample from is, up to an additive constant, given by
\begin{equation*}
\begin{split}
\mathcal{L}  (\mu_j, \xi_j,& \psi_j,\boldsymbol {\tilde s}_{\cdot j}|Z,\boldsymbol\delta,\boldsymbol v, \boldsymbol m) \propto  \\
 & \sum_{t=1}^T  \left[ \ln(c_j^{m_j}\left(\Phi\left(\frac{z_{tj}}{\exp(\frac{s_{tj}}{2})}\right),v_t;\theta_j^{m_j} \right))  + \ln(\varphi\left(\frac{z_{tj}}{\exp(\frac{s_{tj}}{2})}\right)) -\frac{s_{tj}}{2} \right] \\
 & +\ln(\pi_{SV2}(\mu_j, \xi_j, \psi_j,\boldsymbol{ \tilde s}_{\cdot j})),
\end{split}
\end{equation*}
where $\boldsymbol s_{.j}$ is a function of $\boldsymbol {\tilde s_{.j}}$ (see \eqref{eq:partrafo}).
The necessary derivatives of this log posterior are derived (see Appendix \ref{app:svderiv}) for the Leapfrog approximations and then sampling of the marginal parameters is straightforward.

\subsubsection*{Sampling the dependence parameters}
\label{sec:samplecoppar}
The conditional density we need to sample from for the dependence parameters is proportional to
\begin{equation*}
\begin{split}
f(\boldsymbol\delta,\boldsymbol v|Z,\boldsymbol\mu, \boldsymbol\phi, \boldsymbol\sigma, S, \boldsymbol m) &\propto \ell(\boldsymbol\mu, \boldsymbol\phi, \boldsymbol\sigma, S, \boldsymbol\delta,\boldsymbol v, \boldsymbol m|Z) \pi_J(\boldsymbol\mu, \boldsymbol\phi, \boldsymbol\sigma, S, \boldsymbol\delta, \boldsymbol v)\\
& \propto \ell(\boldsymbol\mu, \boldsymbol\phi, \boldsymbol\sigma, S, \boldsymbol\delta,\boldsymbol v, \boldsymbol m|Z) \prod_{j=1}^d\pi_{u}(\delta_j) \\
& \propto \prod_{t=1}^T \prod_{j=1}^d c_j^{m_j}\left(\Phi\left(\frac{z_{tj}}{\exp(\frac{s_{tj}}{2})}\right),v_t;\theta_j^{m_j}\right) \prod_{j=1}^d\pi_{u}(\delta_j).
\end{split}
\end{equation*}
To sample from this density we use the same HMC approach as in Section \ref{sec:onefact}.

\subsubsection*{Sampling the copula family indicators}
The full conditional of $m_j$ is obtained as
\begin{equation*}
f(m_j|Z,\boldsymbol \mu,\boldsymbol \phi,\boldsymbol \sigma ,S,\boldsymbol \delta,\boldsymbol v, \boldsymbol m_{-j}) = \frac{ \prod_{t=1}^T  c_j^{m_j}\left(\Phi\left(\frac{z_{tj}}{\exp(\frac{s_{tj}}{2})}\right),v_t;\theta_j^{m_j}\right)}{\sum_{m_j' \in \mathcal{M}}  \prod_{t=1}^T  c_j^{m_j'}\left(\Phi\left(\frac{z_{tj}}{\exp(\frac{s_{tj}}{2})}\right),v_t;\theta_j^{m_j'}\right) }. 
\end{equation*}
We can sample directly from this discrete distribution and no MCMC updates are required here.

\subsection{Simulation study}
We conduct a simulation study to evaluate the performance of the proposed joint HMC sampler. We consider one scenario in five dimensions and one scenario in ten dimensions, as specified in Table \ref{tab:simscen}. We choose rather high values for the marginal persistence parameter $\phi$ and moderate values for the dependence parameter Kendall's $\tau$. These choices roughly correspond to what we expect to see in financial data. For each scenario we simulate 100 data sets from the model introduced in Section \ref{sec:modelspec}. The proposed MCMC sampler with HMC updates is then applied to the simulated data. The sampler is run for 2500 iterations, whereas the first 500 iterations are discarded for burn in. For family selection we consider the following set of one parametric copula families $\{$Gaussian, Student t(df=4), Clayton, Gumbel$\}$.

\begin{equation}
\begin{split}
\boldsymbol \mu_{sim} &= (-6,-6,-7,-7,-8)\\ 
\boldsymbol \phi_{sim} &= (0.7,0.8,0.85,0.9,0.95)\\ 
\boldsymbol \sigma_{sim} &= (0.2,0.2,0.3,0.3,0.4)\\ 
\boldsymbol \tau_{sim} &= (0.3,0.4,0.5,0.6,0.7)\\ 
\boldsymbol m_{sim} &= (\text{Gaussian, Student t(df=4), Clayton, Gumbel, Gaussian})\\ 
\end{split}
\end{equation}

\begin{table}[H]
\scriptsize
\centering
\begin{tabular}{llllllll}
Scenario & $d$ & $T$ & $\boldsymbol \mu$ & $\boldsymbol \phi$ & $\boldsymbol \sigma$  & $\boldsymbol \tau$ &  $\boldsymbol m$ \\
  \hline 
1 & 5 & 1000 & $\boldsymbol \mu_{sim}$ & $\boldsymbol \phi_{sim}$ & $\boldsymbol \sigma_{sim}$ & $\boldsymbol \tau_{sim}$ & $\boldsymbol m_{sim}$ \\
2 & 10 & 1000 & $(\boldsymbol \mu_{sim},\boldsymbol \mu_{sim})$ & ($\boldsymbol \phi_{sim}$,$\boldsymbol \phi_{sim}$) & ($\boldsymbol \sigma_{sim}$,$\boldsymbol \sigma_{sim}$) & ($\boldsymbol \tau_{sim}$,$\boldsymbol \tau_{sim}$) & $(\boldsymbol m_{sim}, \boldsymbol m_{sim})$  \\
\end{tabular}
\caption{Parameter specification for the two different scenarios in the simulation study.}
\label{tab:simscen}
\end{table}

\begin{table}[H]
\scriptsize
\begin{tabular}{lllllllllll}
  \hline
\textbf{Scenario 1} & $\mu_1$ & $\mu_2$ & $\mu_3$ & $\mu_4$ & $\mu_5$ & $\phi_1$ & $\phi_2$ & $\phi_3$ & $\phi_4$ & $\phi_5$ \\ 
  \hline
MSE & 0.0027 & 0.0038 & 0.0059 & 0.0107 & 0.0851 & 0.0362 & 0.0408 & 0.0059 & 0.0017 & 0.0003 \\ 
C.I. $90\%$ & 0.91 & 0.86 & 0.92 & 0.92 & 0.82 & 0.97 & 0.86 & 0.89 & 0.90 & 0.83  \\ 
C.I. $95\%$& 0.95 & 0.90 & 0.96 & 0.94 & 0.87 & 0.99 & 0.94 & 0.94 & 0.93 & 0.89 \\ 
ESS& 1022 & 666 & 761 & 942 & 505 & 644 & 433 & 399 & 461 & 325 \\ 
   \hline
\end{tabular}
\begin{tabular}{lllllllllll}
  \hline
 & $\sigma_1$ & $\sigma_2$ & $\sigma_3$ & $\sigma_4$ & $\sigma_5$ & $\tau_1$ & $\tau_2$ & $\tau_3$ & $\tau_4$ & $\tau_5$ \\ 
  \hline
MSE & 0.0077 & 0.0055 & 0.0037 & 0.0024 & 0.0026 & 0.0094 & 0.0164 & 0.0255 & 0.0364 & 0.0503  \\ 
 C.I. $90\%$ &0.95 & 0.93 & 0.91 & 0.91 & 0.81 & 0.78 & 0.79 & 0.68 & 0.79 & 0.72 \\ 
C.I. $95\%$ & 0.98 & 0.93 & 0.93 & 0.97 & 0.88 & 0.84 & 0.85 & 0.77 & 0.81 & 0.74  \\ 
ESS& 391 & 368 & 326 & 360 & 255 & 879 & 770 & 528 & 480 & 280 \\ 
   \hline
\end{tabular}
\begin{tabular}{lllllllllll}
  \hline
 & $s_{300,1}$ & $s_{300,2}$ & $s_{300,3}$ & $s_{300,4}$ & $s_{300,5}$ & $v_{100}$& $v_{200}$& $v_{500}$& $v_{800}$& $v_{900}$ \\ 
  \hline
MSE & 0.0564 & 0.0892 & 0.2234 & 0.1836 & 0.2132 & 0.0239 & 0.0241 & 0.0283 & 0.0222 & 0.0202 \\ 
 C.I. $90\%$ & 0.94 & 0.88 & 0.93 & 0.91 & 0.94 & 0.86 & 0.91 & 0.83 & 0.86 & 0.84 \\ 
C.I. $95\%$ & 0.98 & 0.95 & 0.95 & 0.96 & 0.97 & 0.91 & 0.94 & 0.88 & 0.88 & 0.89 \\ 
ESS &  1448 & 433 & 1433 & 1343 & 1334 & 997 & 1104 & 1036 & 1111 & 1085\\ 
   \hline
\end{tabular}
\caption{MSE estimated using the posterior mode, observed coverage probability of the credible intervals (C.I.) and effective samples size calculated from 2000 posterior draws for selected parameters (Scenario 1).}
\label{tab:sumsim1}
\end{table}

The simulation results are summarized in Tables \ref{tab:sumsim1} and \ref{tab:famsel} for the five dimensional scenario and in Tables \ref{tab:sumsim2} and \ref{tab:famsel2} in the appendix for the ten dimensional setup. Comparing these two setups we see that the ESS of the Kendall's $\tau$ parameters and of the latent states $v_t$ is better for the five dimensional scenario.  
Besides that, the results of the five and ten dimensional setups are only slightly different and therefore we discuss only the five dimensional scenario.

\begin{table}[H]
\centering
\begin{tabular}{rrrrrr}
  \hline
 & $m_1$ & $m_2$ & $m_3$ & $m_4$ & $m_5$ \\ 
  \hline
 & 94$\%$ & 90$\%$ & 87$\%$ & 77$\%$ & 66$\%$ \\ 
   \hline
\end{tabular}
\caption{Proportion of how often the correct copula family was selected. The selected copula family is the posterior mode estimate of $m_j$ for $j=1, \ldots, 5$ (Scenario 1).}
\label{tab:famsel}
\end{table}

\begin{figure}[H]
\centering
\includegraphics[width=0.87\textwidth]{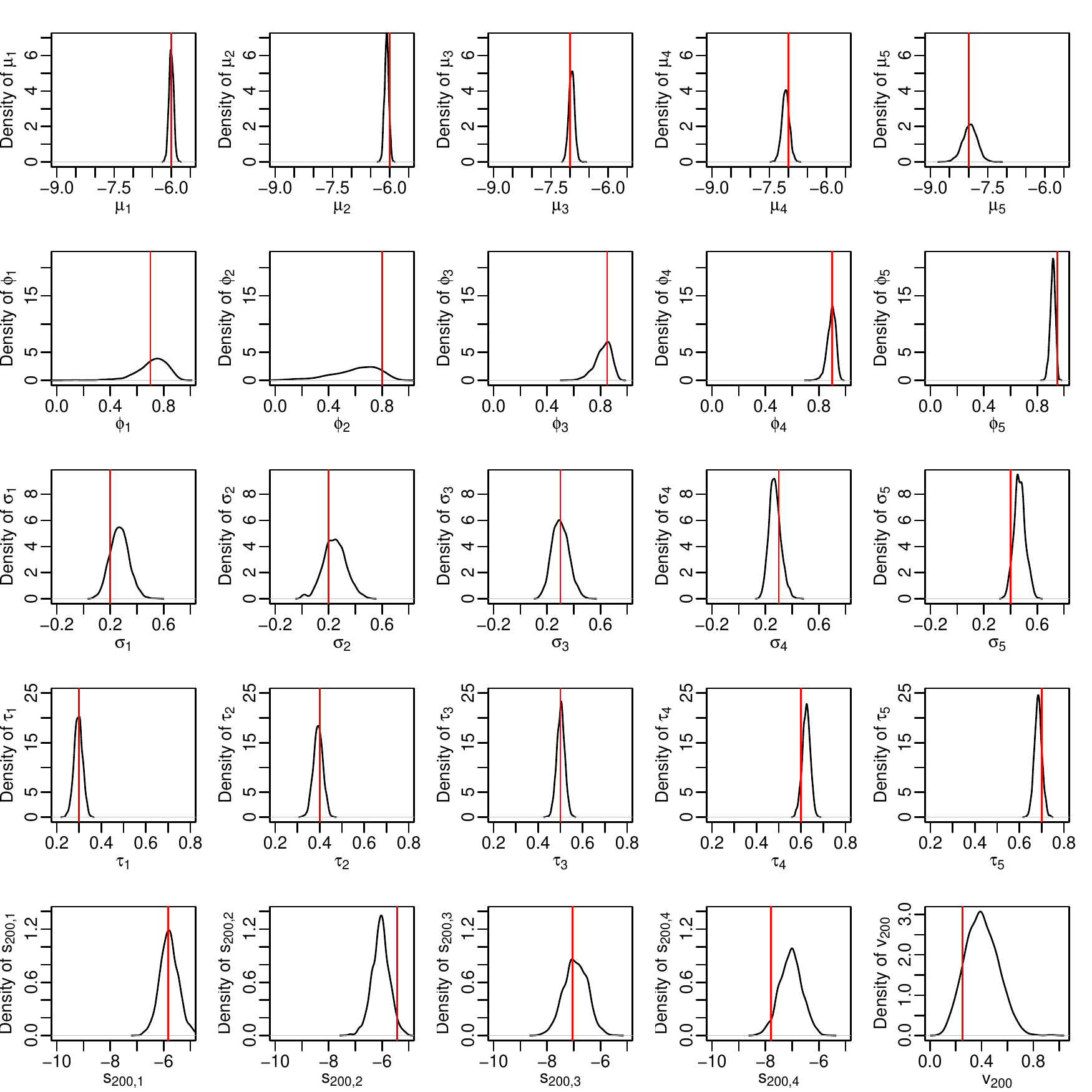}
\caption{Kernel density estimates of the posterior density of selected parameters obtained from a single MCMC run in Scenario 1. The estimates are based on 2000 MCMC iterations after a burn in of 500. The true parameter value is added in red.}
\label{fig:dens}
\end{figure}

\begin{figure}[H]
\centering
\includegraphics[width=0.87\textwidth]{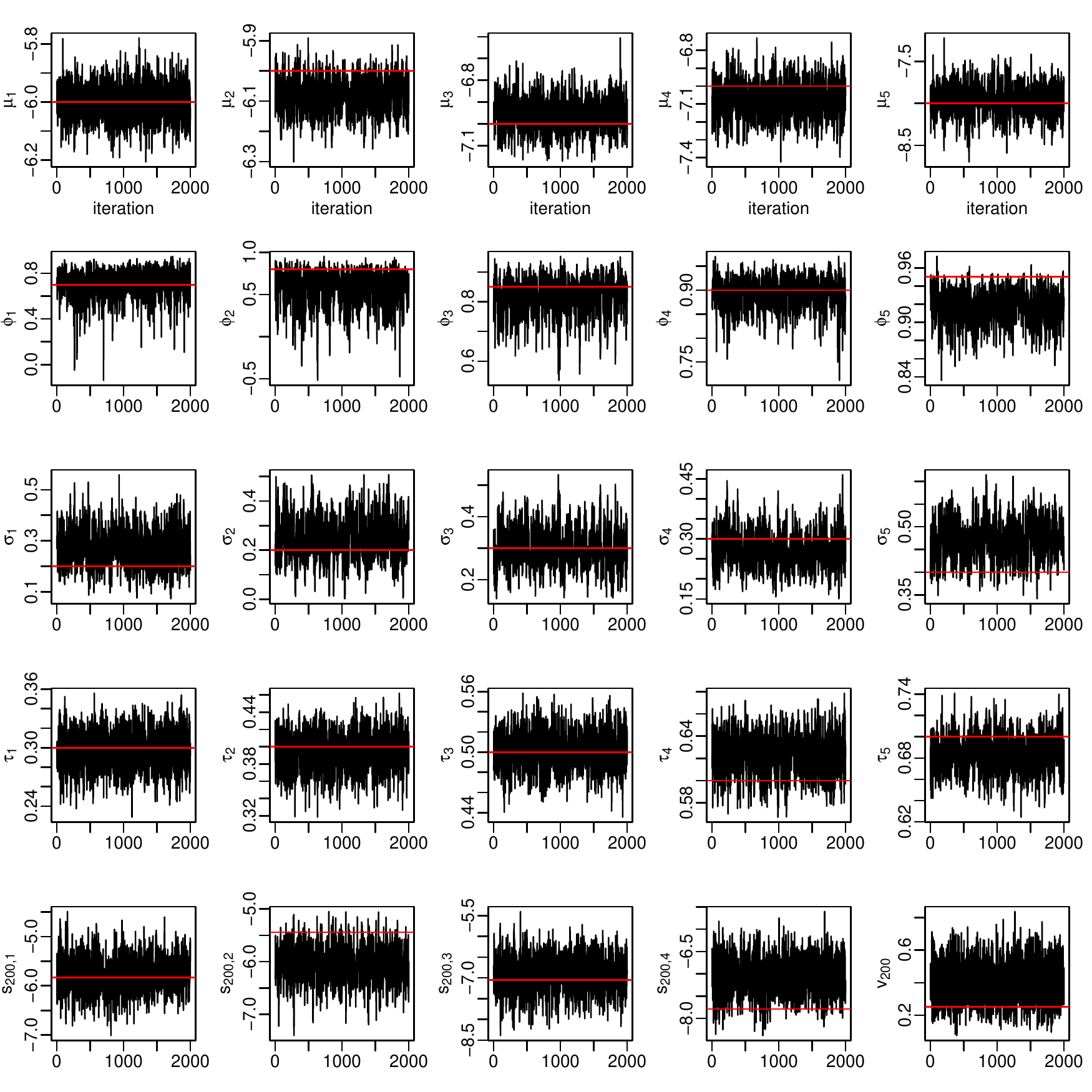}
\caption{Trace plots of selected parameters obtained from a single MCMC run in Scenario 1. The trace plots show 2000 MCMC iterations after a burn in of 500. The true parameter value is added in red.}
\label{fig:trace}
\end{figure}

Comparing the simulation results for the factor copula parameters to the results of Section  \ref{sec:simonefac}, we see that we perform worse in terms of observed coverage probabilities and MSE. But this is not surprising because here we consider a much more complex model and also update the copula families and the marginal parameters. Further we see that the ESS decreases from $\tau_1$ up to $\tau_5$. This is in line with our findings in Section \ref{sec:simonefac} where we have seen that mixing is worse for higher Kendall's $\tau$ values. We can also observe differences with respect to the observed coverage probability of credible intervals. For a low  marginal persistence parameter ($\phi_1$) coverage probabilities are very high suggesting a broad posterior distribution. For a high persistence parameter ($\phi_5$) the observed coverage probabilities are lower.  Figures \ref{fig:dens} and \ref{fig:trace} show estimated posterior densities and trace plots of one MCMC run for the five dimensional setup. These figures suggest that we achieve proper mixing. Furthermore we see that the estimated posterior density of $\phi_1$ is more dispersed compared to the estimated posterior density of $\phi_5$.
Table \ref{tab:famsel} shows that the correct copula family was selected in at least 66 out of 100 cases. This frequency is best for the first linking copula which has a low Kendall's $\tau$ value and worst for the linking copula with the highest Kendall's $\tau$ value.  

Overall, the results suggest that the method performs well.
For all parameters we obtain reasonable MSE and  ESS values and  our method is able to select the correct copula family in most cases.
In particular our HMC schemes do a good job at jointly updating more than $T=1000$ parameters of one Gibbs block.

\section{Application}
\label{sec:app}

We illustrate our approach with one-day ahead value at risk (VaR) prediction for a portfolio consisting of several stocks. These predictions can be obtained from simulations of the predictive distribution.
As before, $Z$ is the data matrix containing $T$ observations of the $d$ stocks. We need to sample from the predictive distribution of the log returns at time $T+1$, $\boldsymbol {Z_{T+1}} = (Z_{T+1,1}, \ldots Z_{T+1,d}) $, given $Z$. We obtain simulations from the joint density
\begin{equation}
f(\boldsymbol {z_{T+1}} ,\boldsymbol s_{T+1\cdot},S,\boldsymbol \mu,\boldsymbol \phi,\boldsymbol \sigma,\boldsymbol \delta,v_{T+1},\boldsymbol v, \boldsymbol m|Z),
\end{equation}
with the following steps:
\begin{itemize}
\item Simulate $S,\boldsymbol\mu,\boldsymbol\phi,\boldsymbol\sigma,\boldsymbol \delta,\boldsymbol v, \boldsymbol m$ from the corresponding posterior distribution given the data $Z$ with our sampler developed in Section \ref{sec:jmod}. We discard the first 500 samples for burnin and denote the remaining $R=2000$ samples by $S^r,\boldsymbol\mu^r,\boldsymbol\phi^r,\boldsymbol\sigma^r,\boldsymbol \delta^r,\boldsymbol v^r, \boldsymbol m^r$, $r=1, \ldots, R$.
\end{itemize}
\hspace*{0.4cm} We proceed as follows for $r=1, \ldots, R$:

\begin{itemize}
\item Simulate $v^r_{T+1} \sim unif(0,1)$.

\item For $j = 1, \ldots, d$ simulate
$
s^r_{T+1,j} \sim N(\mu_j^r + \phi_j^r(s_{Tj}^r - \mu_j^r), (\sigma^r_j)^2).
$

\item To obtain the sample $\boldsymbol z^r_{T+1}$ from
\begin{flalign*}
f(\boldsymbol z_{T+1}| &\boldsymbol s^r_{T+1\cdot},S^r,\boldsymbol\mu^r,\boldsymbol\phi^r,\boldsymbol\sigma^r,\boldsymbol \delta^r,v^r_{T+1},\boldsymbol v^r,\boldsymbol m^r, Z)  = \\ &\prod_{j=1}^d \Bigg[ c^{m^r_j}_j\left(\frac{z_{T+1j}}{\exp(\frac{s^r_{T+1j}}{2})},v^r_{T+1};\theta_j^{m^r_j} \right) \varphi \left(\frac{z_{T+1j}}{\exp(\frac{s^r_{T+1j}}{2})}\right)
 \cdot \frac{1}{\exp(\frac{s^r_{T+1j}}{2})} \Bigg]
\end{flalign*}
we simulate $u_j^{r}$ from $C_j^{m^r_j}\left(\cdot|v^r_{T+1};\theta^{m^r_j}_j\right)$ and set $z^r_{T+1j} = \Phi^{-1}\left(u_j^{r}|0,\exp({s^r_{T+1j}})\right)$ for $j = 1, \ldots, d$. Here $\Phi(\cdot|\mu_{normal}, \sigma_{normal}^2)$ is the distribution function of a normally distributed random variable with mean $\mu_{normal}$ and variance $\sigma_{normal}^2$.

\end{itemize}

\begin{figure}[H]
\centering
\includegraphics[width=
1\textwidth]{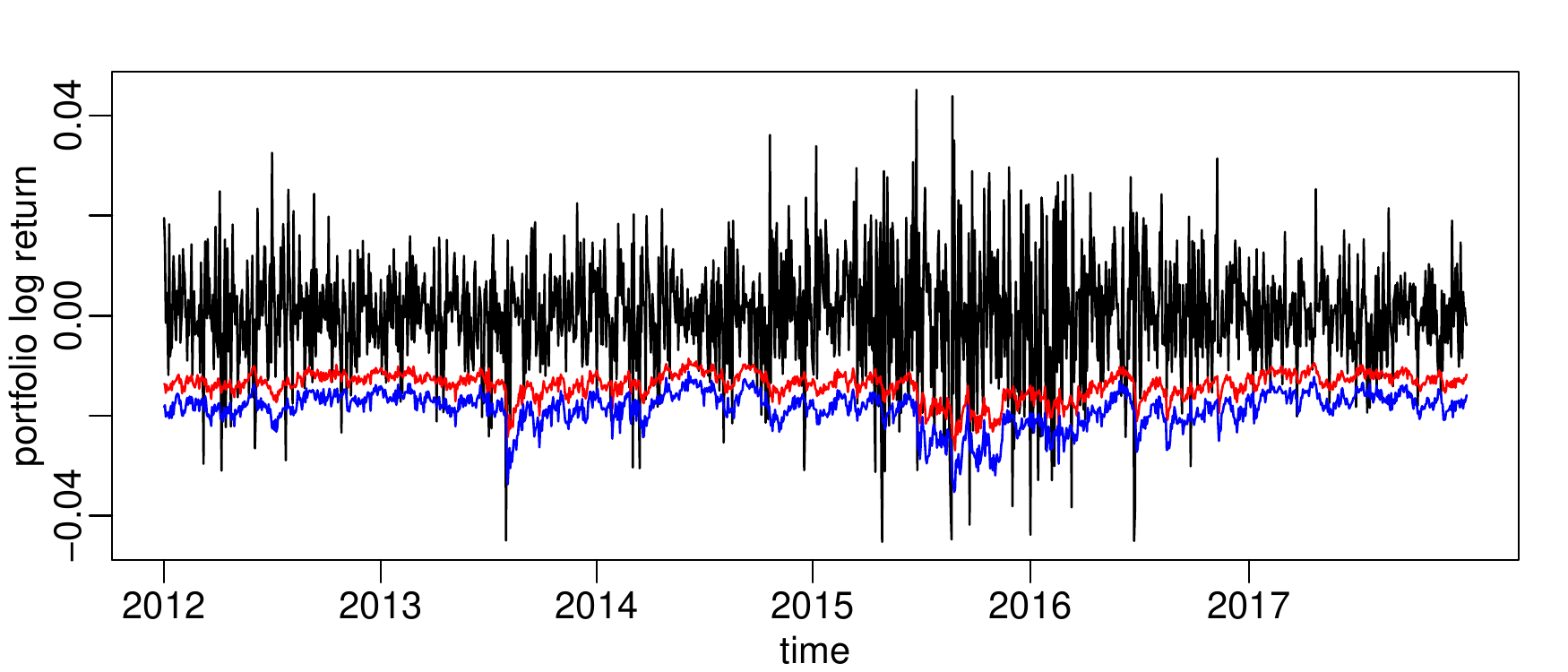}
\caption{Observed daily log return of the portfolio and the estimated one-day ahead $90\%$ VaR (red) and $95\%$ VaR (blue) plotted against time in years.}
\label{fig:var}
\end{figure}

We consider an equally weighted portfolio consisting of 6 stocks from German companies (BASF, Fresenius Medical Care, Fresenius SE, Linde, Merck, K+S). Since all companies are chosen from the chemical/pharmaceutical/medical industry we assume that a model with one factor is suitable to capture the dependence structure. Our data, obtained from Yahoo Finance (https://finance.yahoo.com), contains daily log returns of these stocks from 2008 to 2017. 
We use 1000 days as training period, which corresponds to data of approximately four years. We set $T=1000$ and obtain simulations of the one-day ahead predictive distribution as described above for the first trading day in $2012$. Instead of refitting the model for each day we fix parameters that do not change over time ($\boldsymbol \mu, \boldsymbol \phi, \boldsymbol \sigma, \boldsymbol \delta, \boldsymbol m$) at their posterior mode estimates and only update dynamic parameters ($S, \boldsymbol v$) as in \cite{kreuzer2019efficient} for the remaining one-day ahead predicitve simulations. For updating only the dynamic parameters we have seen that it is enough to use the last $100$ time points and time needed for computation is reduced a lot. We obtain $2000$ simulations of the one-day ahead predictive distribution for each trading day in the period from January 2012 to December 2017. From the simulations we calculate the portfolio value, and take the corresponding quantile to obtain the VaR prediction. We consider the same VaR level of $90\%$ as in \cite{schamberger2017bayesian} and additionally the $95\%$ VaR.
The linking copulas are chosen from the following set of one parametric copula families: Gaussian, Student t with 4 degrees of freedom, (survival) Gumbel and (survival) Clayton. For a copula with density $c(u_1,u_2)$, the corresponding survival copula has density $c(1-u_1,1-u_2)$. More details about the considered bivariate copula families are given in \cite{czadoanalyzing}, Chapter 3. With these choices we cover a range of different tail dependence structures: no tail dependence (Gauss), symmetric tail dependence (Student t) and asymmetric tail dependence ((survival) Gumbel and (survival) Clayton). As explained above, the copula family indicator was only updated for the first model we fitted and then kept fixed. The linking copula families of BASF, Fresenius Medical Care, Fresenius SE, Linde, Merck and K+S with the highest posterior probabilities are Student t, survival Gumbel, survival Gumbel, Gaussian, survival Gumbel and Student t, respectively. In particular, we obtain a model with asymmetric tail dependence structure. Predicting the VaR for each trading day in six years results in 1521 VaR predictions. The portfolio log returns and corresponding $90\%$ and $95\%$ VaR predictions are visualized in Figure \ref{fig:var}. We observe that the one-day ahead VaR forecast adapts to changes in the volatility. 

To benchmark the proposed model (factor copula SV (\textbf{fc SV})) we repeated the procedure for VaR prediction with two other models: marginal dynamic linear models combined with single factor copulas (\textbf{fc  dlm}) estimated with a two-step procedure as proposed by \cite{schamberger2017bayesian} and a multivariate factor stochastic volatility model with dynamic factors (\textbf{df Gauss SV}) as proposed by \cite{kastner2017efficient}. The df Gauss SV model is here restricted to one factor. To illustrate the necessity of copula family selection we further consider \textbf{fc SV} models with the restriction that all linking copula are chosen from the same family. We consider the three copula families that were selected as linking copulas for the \textbf{fc SV} model and obtain the three restricted models \textbf{fc SV (Ga)}, \textbf{fc SV (t)} and \textbf{fc SV (sGu)} which have only Gaussian, Student t(df=4) and survival Gumbel linking copulas, respectively. 
Additionally we compare the proposed approach to a two-step estimation of the factor copula SV model (\textbf{fc SV (ts)}). In this two-step approach we obtain simulations from the predictive distribution of the log returns at time $T+1$, $\boldsymbol {Z_{T+1}} $, given $Z$ as follows:

\begin{itemize}
\item Estimate a SV model for each margin separately and obtain marginal posterior mode estimates for the latent log variances denoted by $\hat s_{tj}$ for $t=1, \ldots, T, j=1, \ldots, d$. 
\item Use the probability integral transform to obtain data on the (0,1) scale,\newline $u_{tj} \coloneqq \Phi\left(z_{tj} \cdot \exp(-\frac{\hat s_{tj}}{2})\right)$.
\item  For the data $u_{tj}, t=1, \ldots, T, j=1, \ldots, d$, we fit the single factor copula model with HMC as explained in Section \ref{sec:onefact}, where we allow for Bayesian copula family selection and obtain posterior mode estimates of the corresponding parameters denoted by $\hat\delta_1, \ldots \hat\delta_d, \hat m_1, \ldots \hat m_d$.
\item For each margin, we simulate from the predictive distribution of the log variances at time $T+1$, i.e. from $s_{T+1,j}|z_{1j}, \ldots, z_{Tj}$, and obtain marginal posterior mode estimates $\hat s_{T+1,j}$ for $j=1, \ldots, d$.
\end{itemize}
\hspace*{0.4cm} For $r=1, \ldots, R$ we proceed as follows:
\begin{itemize}
\item We simulate $u_{1}^r, \ldots u_{d}^{r}$ from the single factor copula with parameters $\hat\delta_1, \ldots \hat\delta_d, \hat m_1, \ldots \hat m_d$.
\item We set $z_{T+1,j}^r = \Phi^{-1}\left(u_{j}^{r}|0,\exp({\hat s_{T+1,j}})\right)$ for $j=1, \ldots, d$.
\end{itemize}

Standard measures to compare the predictive accuracy between different models are the continuous ranked probability score (\cite{gneiting2007strictly}) or log predictive scores as used in \cite{kastner2016sparse}. These scores evaluate the overall performance. But we are interested in the VaR, a quantile of the predictive distribution, which is only one specific aspect. Therefore we use the rate of VaR violations and Christoffersen's conditional coverage test (\cite{christoffersen2012elements}, Chapter 8), which are commonly used to compare VaR forecasts, as in \cite{schamberger2017bayesian} and \cite{nagler2019model}.  
 From an optimal VaR measure at level $p$ we would expect that there are $(1-p) \cdot 100 \%$ VaR violations and that violations occur independently. This constitutes the null hypotheses of Christoffersen's conditional coverage test.
The VaR violation rates for the different models are shown in Table \ref{tab:freq}. For the $90\%$ VaR, the violation rate of the df Gauss SV model is closest to the optimal rate of $10\%$, whereas for the $95\%$ VaR the fc SV model performs best. According to the p-values of Christoffersen's conditional coverage test in Table \ref{tab:pval} none of the considered  models can be rejected at the $5\%$ or $10\%$ level with respect to $90\%$ VaR prediction. But with respect to $95\%$ VaR prediction, every model except the fc SV model is rejected at the $5\%$ or $10\%$ level. We conclude that the preferred model in this scenario is the fc SV model.

\begin{table}[ht]
\scriptsize
\centering
\begin{tabular}{lrrrrrrrr}
  &\multicolumn{1}{c}{\textbf{fc SV}} & \multicolumn{1}{c}{\textbf{fc SV (Gauss)
  }}& \multicolumn{1}{c}{\textbf{fc SV (t)
  }}& \multicolumn{1}{c}{\textbf{fc SV (sGu)
  }} & \multicolumn{1}{c}{\textbf{fc SV (ts)}} & \multicolumn{1}{c}{\textbf{fc dlm}} & \multicolumn{1}{c}{\textbf{df Gauss SV}}\\
 \hline
$90\%$ VaR  & $9.07\%$  & 8.74   &  9.27 & 8.68 &$10.52 \%$ & $8.74\%$&$\boldsymbol{10.32} \%$\\
$95\%$ VaR & $\boldsymbol{4.93\%}$  & 5.39 &5.26& 4.14  &  $6.51 \%$&$ 4.80 \%$ & $5.85\%$ \\
   \hline
\end{tabular}
\caption{The rate of $90\%$ and $95\%$ VaR violations for the seven models: fc SV, fc SV(Gauss), fc SV(t), fc SV(sGu), fc SV(ts), fc dlm, df Gauss SV. The violation rate closest to the opimal value of $5\%$ or $10 \%$ is marked in bold.}
\label{tab:freq}
\end{table}

\begin{table}[ht]
\scriptsize
\centering
\begin{tabular}{lllllllll}
 &\multicolumn{1}{c}{\textbf{fc SV}} & \multicolumn{1}{c}{\textbf{fc SV (Gauss)
  }}& \multicolumn{1}{c}{\textbf{fc SV (t)
  }}& \multicolumn{1}{c}{\textbf{fc SV (sGu)
  }} & \multicolumn{1}{c}{\textbf{fc SV (ts)}} & \multicolumn{1}{c}{\textbf{fc dlm}} & \multicolumn{1}{c}{\textbf{df Gauss SV}}\\
 \hline
$90\%$ VaR & $0.13$  &  0.1  &0.43  & 0.05  &$\boldsymbol{ 0.78}$ & 0.19&  $0.69 $ \\
$95\%$ VaR & $\boldsymbol{0.12}$  &0.04  & 0.03&0 & 0.01 & 0.03 &0.02 \\
   \hline
\end{tabular}
\caption{The p-value of Christoffersen's conditional coverage test for the $90\%$ and $95\%$ VaR predictions of seven models: fc SV, fc SV(Gauss), fc SV(t), fc SV(sGu), fc SV(ts), fc dlm, df Gauss SV. The highest p-value per row is marked in bold.}
\label{tab:pval}
\end{table}

\section{Conclusion}
\label{sec:conc}
We propose a single factor copula SV model, a combination of the SV model for the margins and factor copulas for the dependence. Dependence and marginal parameters are estimated jointly within a Bayesian approach, avoiding a two-step estimation procedure which is commonly used for copula models. The proposed model can be seen as one way to extend factor SV models that rely on Gaussian dependence to more complex dependence structures. 
The necessity of such models was illustrated with one-day ahead value at risk prediction. In the application our stocks were chosen such that one factor is suitable to describe dependencies. However this might not be appropriate for different portfolios and the extension of the proposed model to multiple factors will be subject to future research. This extension to multiple factors could exploit the partition of different stocks into sectors as in the structured factor copula model proposed by \cite{krupskii2015structured}. Another extension could allow for time varying dependence parameters or for copula families with two and more parameters.

\section*{Acknowledgment}

The first author acknowledges financial support by a research stipend of the Technical University of Munich. The second author is supported by  the  German  Research  Foundation  (DFG grant
CZ 86/4-1). Computations were performed on a Linux cluster supported by DFG grant INST 95/919-1 FUGG.

\bibliographystyle{spbasic}    
\bibliography{References}{}

\section{Appendix}

\subsection{Hamiltonian Monte Carlo}
We provide a short introduction to HMC based on \cite{neal2011mcmc}. We start with the introduction of the Hamiltonian dynamics. 
\subsubsection*{Hamiltonian dynamics}
We consider a position vector $\boldsymbol q \in \mathbb{R}^d$ with associated momentum vector $\boldsymbol p \in \mathbb{R}^d$ at time $t$. Their change over time is described through the function $H(\boldsymbol p,\boldsymbol q)$, the \emph{Hamiltonian}, which satisfies the following differential equations:
\begin{equation}
\begin{split}
\frac{dq_i}{dt}&=\frac{dH}{dp_i} \\
\frac{dp_i}{dt}&=-\frac{dH}{dq_i}, i=1, \ldots, d.
\end{split}
\label{eq:hdyn}
\end{equation}
Here $H$ represents the total energy of the system.
In HMC, it is assumed that $H$ can be expressed as
\begin{equation}
H(\boldsymbol p,\boldsymbol q) = U(\boldsymbol q) + K(\boldsymbol p) = U(\boldsymbol q) + \boldsymbol p^\top M^{-1}\boldsymbol p/2,
\label{eq:hform}
\end{equation}
where $U(\boldsymbol q)$ is called the potential energy and $K(\boldsymbol p)$ the kinetic energy. Further $M$ is a symmetric positive definite mass matrix, which is usually assumed to be diagonal.
The Hamiltonian dynamics, specified in \eqref{eq:hdyn}, can therefore be rewritten as
\begin{equation}
\begin{split}
\frac{dq_i}{dt}&=(M^{-1}\boldsymbol p)_i \\
\frac{dp_i}{dt}&=-\frac{dU}{dq_i}, i=1, \ldots, d.
\end{split}
\label{eq:hdyn2}
\end{equation}
\subsubsection*{Leapfrog method}
Since it is usually not possible to solve the system of differential equations given in \eqref{eq:hdyn2} analytically we need to find iterative approximations. Therefore we use the Leapfrog method, where the state one-step ahead of time $t$ with step size $\epsilon$, i.e. the state at time $t+\epsilon$, is approximated by
\begin{equation*}
\begin{split}
p_i(t+\epsilon/2) &= p_i(t) - \frac{\epsilon}{2} \frac{dU}{dq_i}(\boldsymbol q(t)) \\
q_i(t+\epsilon) &= q_i(t) + \epsilon \frac{p_i(t+\epsilon /2)}{m_i} \\
p_i(t+\epsilon) &= p_i(t+\epsilon/2) - \frac{\epsilon}{2} \frac{dU}{dq_i}(\boldsymbol q(t+\epsilon)), \text{ for } i = 1, \ldots d. \\
\end{split}
\end{equation*}

\subsubsection*{Canonical distribution}

To use Hamiltonian dynamics within MCMC sampling we need to relate the energy function to a probability distribution. Therefore we utilize the \emph{canonical distribution} $P(\boldsymbol x)$ associated with a general energy function $E(\boldsymbol x)$ with state $\boldsymbol x$ defined through the density 
\begin{equation*}
p(\boldsymbol x) \coloneqq \frac{1}{Z} \exp(-E(\boldsymbol x)/T).
\end{equation*}
Here $T$ is the temperature of the system and $Z$ the normalizing constant needed to satisfy the density constraint.
So the Hamiltonian $H(\boldsymbol p,\boldsymbol q)$ specified in \eqref{eq:hform} defines a probability density given by
\begin{equation*}
p(\boldsymbol q,\boldsymbol p) = \frac{1}{Z} \exp(-H(\boldsymbol p,\boldsymbol q)/T) = \frac{1}{Z} \exp(-U(\boldsymbol q)/T)\exp(-K(\boldsymbol p)/T),
\end{equation*}
where $\boldsymbol q$ and $\boldsymbol p$ are independent. In the following we assume $T=1$.

\subsubsection*{Sampling with Hamiltonian Monte Carlo}

In HMC we specify the corresponding energy function of $\boldsymbol q$ and $\boldsymbol p$, i.e. the Hamiltonian, and sample from the corresponding canonical distribution of $\boldsymbol q$ and $\boldsymbol p$. In a Bayesian setup we identify $\boldsymbol q$ as our parameters of interest and $\boldsymbol p$ are auxiliary variables. Therefore we set 
\begin{equation*}
U(\boldsymbol q) \coloneqq -\ln(\pi(\boldsymbol q)\ell(\boldsymbol q|D)),
\end{equation*}
where $\pi(\boldsymbol q)$ is the prior density and $\ell(\boldsymbol q|D)$ the likelihood function for the given data $D$. Therefore the canonical distribution of $\boldsymbol q$ corresponds to the posterior distribution of $\boldsymbol q$, when $T=1$.

 Since $K(\boldsymbol p) =  \boldsymbol p^\top M^{-1}\boldsymbol p/2$, it holds that the auxiliary parameter vector $\boldsymbol p$ is multivariate normal distributed with zero mean vector and covariance matrix $M$. Sampling is then done in the following way.
\begin{enumerate}
\item Sample $\boldsymbol p$ from the normal distribution with zero mean vector and covariance matrix $M$.
\item Metropolis update: Start with the current state $(\boldsymbol q,\boldsymbol p)$ and simulate $L$ steps of Hamiltonian dynamics  with step size $\epsilon$ using the Leapfrog method. Obtain $(\boldsymbol q',\boldsymbol p')$ and accept this proposal with Metropolis acceptance probability
\begin{equation*}
\min(1,\exp(-H(\boldsymbol q',\boldsymbol p')+H(\boldsymbol q,\boldsymbol p))) = \min\left(1,\frac{\pi(\boldsymbol q')l(\boldsymbol q'|D)     \exp(\boldsymbol p^\top M^{-1}\boldsymbol p/2)}{\pi(\boldsymbol q)l(\boldsymbol q|D)    \exp(\boldsymbol p'^\top M^{-1}\boldsymbol p'/2)}\right).
\end{equation*}
\end{enumerate}

\subsection{Derivatives for HMC for the single factor copula model}
\label{seq:deriv_hmc}
The derivatives of the log posterior density with respect to the parameters $\delta_{j'}$ and $w_{t'}$ are given by
\begin{equation*}
\begin{split}
\dv{}{\delta_{j'}}\mathcal{L}(\boldsymbol\delta_{1:d}, \boldsymbol w_{1:T}|U_{1:T,1:d}) =& \sum_{j=1}^d \sum_{t=1}^T \dv{}{\delta_{j'}} \ln(c_j(u_{tj},v_t;\theta_{j}))
 +\dv{}{\delta_{j'}}\ln(\pi_{FC}(\boldsymbol\delta_{1:d},\boldsymbol w_{1:T})) \\
=& \sum_{t=1}^T \dv{}{\delta_{j'}} \ln(c_j(u_{tj'},v_t;\theta_{j'})) +\dv{}{\delta_{j'}}\ln(\pi_{FC}(\boldsymbol\delta_{1:d},\boldsymbol w_{1:T})) \\
=& \sum_{t=1}^T \dv{}{\theta_{j'}} \ln(c_j(u_{tj'},v_t;\theta_{j'})) \dv{\theta_{j'}}{\delta_{j'}} +\dv{}{\delta_{j'}}\ln(\pi_{FC}(\boldsymbol\delta_{1:d},\boldsymbol w_{1:T})),\\
\end{split}
\end{equation*}
and

\begin{equation*}
\begin{split}
\dv{}{w_{t'}}\mathcal{L}(\boldsymbol\delta_{1:d},\boldsymbol w_{1:T}|U_{1:T,1:d}) =& \sum_{j=1}^d \sum_{t=1}^T \dv{}{w_{t'}} \ln(c_j(u_{tj},v_t;\theta_{j})) +\dv{}{w_{t'}}\ln(\pi_{FC}(\boldsymbol\delta_{1:d},\boldsymbol w_{1:T}))\\
=& \sum_{j=1}^d \dv{}{w_{t'}} \ln(c_j(u_{t'j},v_{t'};\theta_{j})) +\dv{}{w_{t'}}\ln(\pi_{FC}(\boldsymbol\delta_{1:d},\boldsymbol w_{1:T}))\\
=& \sum_{j=1}^d \dv{}{v_{t'}} \ln(c_j(u_{t'j},v_{t'};\theta_{j})) \dv{v_{t'}}{w_{t'}} +\dv{}{w_{t'}}\ln(\pi_{FC}(\boldsymbol\delta_{1:d},\boldsymbol w_{1:T})) \\,
\end{split}
\end{equation*}
The components of the derivative of the log posterior density are derived in the following.

\subsubsection*{Derivatives of the log prior density}
The derivative of the log prior density $\pi_u$ is given by 
\begin{equation*}
\dv{}{x}\ln(\pi_u(x)) = \dv{}{x} -2 \ln(1+\exp(-x)) - x = 2 (1+\exp(x))^{-1} - 1.
\end{equation*}

\subsubsection*{Derivatives of the parameter transformation}

We consider derivatives of the parameter transformation, i.e. $\dv{\theta_{j'}}{\delta_{j'}}$ and  $\dv{v_{t'}}{w_{t'}}$. In this part we suppress
 the indices $j'$ and $t'$. 
We have that
\begin{equation*}
v = (1 + \exp(-w))^{-1},
\end{equation*}
and the derivative is given by
\begin{equation*}
\dv{v}{w} = (1 + \exp(-w))^{-2} \exp(-w).
\end{equation*}

Now we address the derivative $\dv{\theta}{\delta}$. The parameter $\delta$ was chosen to be the logit transform  of the corresponding Kendall's $\tau$ and so $\tau$ can be written as 
\begin{equation*}
\tau = (1 + \exp(-\delta))^{-1},
\end{equation*}
with corresponding derivative
\begin{equation*}
\dv{\tau}{\delta} = (1 + \exp(-\delta))^{-2} \exp(-\delta).
\end{equation*}
The copula parameter $\theta$ is a function of Kendall's $\tau$ ($\theta = g^{-1} (\tau)$) and dependent on the copula family considered we obtain the following derivatives.
\begin{itemize}

\item Gauss and Student t copula
\begin{equation*}
\begin{split}
\theta&=\sin(\frac{1}{2}\pi\tau)\\
\dv{\theta}{\delta} &= \dv{\theta}{\tau} \dv{\tau}{\delta}\\
& =\frac{1}{2}\pi\cos(\frac{1}{2}\pi\tau)\dv{\tau}{\delta}\\
& = \frac{1}{2}\pi\cos(\frac{1}{2}\pi(1 + \exp(-\delta))^{-1})  (1 + \exp(-\delta))^{-2} \exp(-\delta)
\end{split}
\end{equation*}

\item Clayton copula
\begin{equation*}
\begin{split}
\theta&=\frac{2\tau}{1-\tau}\\
 &=\frac{2}{\tau^{-1}-1}\\
&=\frac{2}{1+\exp(-\delta)-1}\\
&=2 \exp(\delta) \\
 \dv{\theta}{\delta}&= 2 \exp(\delta)
\end{split}
\end{equation*}
\item Gumbel copula
\begin{equation*}
\begin{split}
\theta&=(1-\tau)^{-1}\\
\dv{\theta}{\delta} &= \dv{\theta}{\tau} \dv{\tau}{\delta}\\
& = (1-\tau)^{-2} \dv{\tau}{\delta}\\
& = \{1-[1+\exp(-\delta)]^{-1}\}^{-2} [1+\exp(-\delta)]^{-2} \exp(-\delta)\\
&=[1+\exp(-\delta)-1]^{-2} \exp(-\delta)\\
&=\exp(-\delta)^{-2} \exp(-\delta)\\
&=\exp(\delta)
\end{split}
\end{equation*}
\end{itemize}

\subsubsection*{Derivatives of log copula densities}
 For all considered copula families \cite{schepsmeier2014derivatives} calculate the derivatives of the copula density with respect to the copula parameter $\theta_j$ and with respect to the argument $v_t$. Based on their results the derivatives of the log copula density are easily derived.
The derivatives are also implemented in the $\texttt{R}$ package $\texttt{VineCopula}$ by \cite{schepsmeier2018package}.

\subsection{Prior densities for transformed parameters}
\label{sec:trafo_dens}
The prior densities for $\mu_j,\phi_j$ and $\sigma^2_j$ in \eqref{eq:priorfull} imply the following prior densities for $\mu_j,\xi_j$ and $\psi_j$.

\begin{itemize}
\item We have that $\mu_j \sim N(0,\sigma_{\mu}^2) $. So the prior density for $\mu_j$ is up to a constant given by
\begin{equation*}
\pi_{\mu}(x) \propto \exp(-\frac{x^2}{2\sigma_{\mu}^2}).
\end{equation*}

\item We have that $\frac{\phi_j+1}{2} \sim Beta(a,b)$. So the density of $\phi_j$ is given by
\begin{equation*}
f_{\phi}(x)=f_{Beta}\left(\frac{x+1}{2}\right) \frac{1}{2}.
\end{equation*}
This implies that the prior density of $\xi_j$ is
\begin{equation*}
\begin{split}
\pi_{\xi}(x)&=f_{\phi}(F_Z^{-1}(x)) \abs{\dv{}{x}F_Z^{-1}(x)}\\
&=\frac{\Gamma(a+b)}{\Gamma(a)\Gamma(b)} \left(\frac{F_Z^{-1}(x) + 1}{2}\right)^{a-1} \left(1-\frac{F_Z^{-1}(x) + 1}{2}\right)^{b-1}\frac{1}{2}\left(1-(F_Z^{-1}(x))^2 \right).
\end{split}
\end{equation*}

\item We have that $\sigma_j^2 \sim \chi^2_1$, i.e.
\begin{equation*}
f_{\sigma}(x) = 2 x  f_{\chi^2_1} (x^2) .
\end{equation*}
So the prior density for $\psi_j$ is given by
\begin{equation*}
\begin{split}
\pi_{\psi}(x) &= f_{\sigma} (\exp(x)) \exp(x) \\
&= 2 \exp(x) \exp(x) \frac{1}{\sqrt{2}\Gamma(\frac{1}{2})} \exp(-x) \exp(-\frac{\exp(2x)}{2}) \\
&=2 \frac{1}{\sqrt{2}\Gamma(\frac{1}{2})} \exp(x) \exp(-\frac{\exp(2x)}{2})   .
\end{split}
\end{equation*}
\end{itemize}

\subsection{Derivatives for HMC for the stochastic volatility model}
\label{app:svderiv}
We need to calculate derivatives of the function
\begin{equation*}
\begin{split}
\mathcal{L}(\mu_j, \xi_j, \psi_j,\boldsymbol {\tilde s}_{\cdot j}|Z,\boldsymbol\delta,\boldsymbol v, \boldsymbol m) \propto & \sum_{t=1}^T \left[ \ln(c_j^{m_j}\left(\Phi\left(\frac{Z_{tj}}{\exp(\frac{s_{tj}}{2})}\right),v_t;\theta_j^{m_j} \right))  + \ln(\varphi\left(\frac{Z_{tj}}{\exp(\frac{s_{tj}}{2})}\right)) -\frac{s_{tj}}{2} \right]\\
 & +\ln(\pi_{SV2}(\mu_j, \xi_j, \psi_j,\boldsymbol {\tilde s}_{\cdot j})),
\end{split}
\end{equation*}
where $\propto$ refers to proportionality up to an additive constant.
To shorten notation we omit the index $j$ in the following and consider the function

\begin{equation*}
\begin{split}
\mathcal{L}_2(\mu, \xi, \psi,\boldsymbol {\tilde s}_{0:T}|Z,\boldsymbol\delta, \boldsymbol v, \boldsymbol m) = & \sum_{t=1}^T \left[ \ln(c\left(\Phi\left(\frac{Z_{t}}{\exp(\frac{s_{t}}{2})}\right),v_t;\theta \right))  + \ln(\varphi\left(\frac{Z_{t}}{\exp(\frac{s_{t}}{2})}\right)) -\frac{s_{t}}{2} \right]\\
 & +\ln(\pi_{SV2}(\mu, \xi, \psi,\boldsymbol {\tilde s}_{0:T})).
\end{split}
\end{equation*}

We define
\begin{equation*}
\Omega(\boldsymbol s_{1:T}) = \sum_{t=1}^T \left( \ln(c\left(\Phi\left(\frac{Z_{t}}{\exp(\frac{s_{t}}{2})}\right),v_t;\theta\right))  + \ln(\varphi\left(\frac{Z_{t}}{\exp(\frac{s_{t}}{2})}\right)) -\frac{s_{t}}{2} \right),
\end{equation*}

and the derivatives can be expressed as
\begin{itemize}
\item
$
\dv{}{\mu} \mathcal{L}_2(\mu, \xi, \psi,\boldsymbol {\tilde s}_{0:T}) = \dv {\Omega(\boldsymbol s_{1:T})}{\boldsymbol s_{1:T}}
\dv{\boldsymbol s_{1:T}}{\mu}
+\dv{}{\mu}\ln(\pi_{SV2}(\mu, \xi, \psi,\boldsymbol {\tilde s}_{0:T}))
$

\item
$
\dv{}{\xi} \mathcal{L}_2(\mu, \xi, \psi,\boldsymbol {\tilde s}_{0:T}) = \dv {\Omega(\boldsymbol s_{1:T})}{\boldsymbol s_{1:T}}
\dv{\boldsymbol s_{1:T}}{\phi}
\dv{\phi}{\xi} +\dv{}{\xi}\ln(\pi_{SV2}(\mu, \xi, \psi,\boldsymbol {\tilde s}_{0:T}))
$

\item
$
\dv{}{\psi} \mathcal{L}_2(\mu, \xi, \psi,\boldsymbol {\tilde s}_{0:T}) = \dv {\Omega(\boldsymbol s_{1:T})}{\boldsymbol s_{1:T}}
\dv{\boldsymbol s_{1:T}}{\sigma}
\dv{\sigma}{\psi}+\dv{}{\psi}\ln(\pi_{SV2}(\mu, \xi, \psi,\boldsymbol {\tilde s}_{0:T}))
$

\item
$
\dv{}{\boldsymbol {\tilde s}_{0:T}} \mathcal{L}_2(\mu, \xi, \psi,\boldsymbol {\tilde s}_{0:T}) = \dv {\Omega(\boldsymbol s_{1:T})}{\boldsymbol s_{0:T}} J+\dv{}{\boldsymbol {\tilde s}_{0:T}}\ln(\pi_{SV2}(\mu, \xi, \psi,\boldsymbol {\tilde s}_{0:T})),
$
\end{itemize}
where $J \in \mathbb{R}^{(T+1) \times (T+1)}$ denotes the corresponding Jacobian matrix, i.e.
$
J_{tj} = \dv{s_t}{\tilde s_j}.
$
The derivatives are calculated in the following.
\begin{itemize}
\item
$
\dv{}{s_i} \Omega(\boldsymbol s_{1:T}) = \dv{}{x} \ln(c(x,v_i;\theta))\Big|_{x=\Phi\left(\frac{Z_{i}}{\exp(\frac{s_{i}}{2})}\right)} \varphi\left(\frac{Z_{i}}{\exp(\frac{s_{i}}{2})}\right)  \frac{Z_{i}}{\exp(\frac{s_{i}}{2})} (-\frac{1}{2}) + \frac{Z_i^2}{2\exp(s_i)} - \frac{1}{2} \text{ for }i=1, \ldots, T
$
\item
We have that
$
s_0 = \frac{\tilde s_0 \sigma}{\sqrt{1-\phi^2}} + \mu, ~~~ s_t = \tilde s_t \sigma + \mu + \phi(s_{t-1} - \mu), t=1, \ldots, T\\
$
and obtain
\begin{table}[H]
\center
\begin{tabular}{ll}
$\dv{s_0}{\mu} = 1$ & $\dv{s_t}{\mu} = 1 - \phi + \phi \dv{}{\mu}s_{t-1} , t=1, \ldots, T$\\
$\dv{s_0}{\phi} = \tilde s_0 \sigma (1-\phi^2)^{-\frac{3}{2}} \phi$ & $\dv{s_t}{\phi} = s_{t-1}-\mu + \phi \dv{}{\phi}s_{t-1}, t=1, \ldots, T$\\
$\dv{s_0}{\sigma} = \frac{\tilde s_0}{\sqrt{1-\phi^2}}$ & $\dv{s_t}{\sigma} = \tilde s_t + \phi \dv{}{\sigma} s_{t-1} , t=1, \ldots, T $\\

$\dv{s_t}{\tilde s_0} = \phi^{t} \frac{\sigma}{\sqrt{1-\phi^2}}, t=0,\ldots, T$ & $\dv{s_t}{\tilde s_j} = \phi^{t-j} \sigma \mathbbm{1}_{t\geq j}  t=0,\ldots, T, j=1, \ldots T$\\

\end{tabular}
\end{table}

\item
$
\dv{\phi}{\xi} = 1 - F^{-1}(\xi)^2, ~~~
\dv{\sigma}{\psi} = \exp(\psi)
$

\item
$
\dv{}{\mu} \ln(\pi_{SV2}(\mu, \xi, \psi,\boldsymbol {\tilde s}_{0:T})) = -\frac{\mu^2}{\sigma_{\mu}^2}
$

\item
$
\dv{}{\xi}\ln(\pi_{SV2}(\mu, \xi, \psi,\boldsymbol {\tilde s}_{0:T}))=(a-1)\frac{(1-F_Z^{-1}(\xi)^2)}{(F_Z^{-1}(\xi)+1)} - (b-1)(1+F_Z^{-1}(\xi)) - 2  F_Z^{-1}(\xi) 
$
where $a=5$ and $b=1.5$ are the parameters of the beta distribution.
\item $\dv{}{\psi}\ln(\pi_{SV2}(\mu, \xi, \psi,\boldsymbol {\tilde s}_{0:T})) = 1 - \exp(2\psi)$

\item $\dv{}{\boldsymbol \tilde s_{0:T}}\ln(\pi_{SV2}(\mu, \xi, \psi,\boldsymbol {\tilde s}_{0:T})) = -\boldsymbol {\tilde s}_{0:T}$
\end{itemize}

\subsection{Results of the simulation study for $d=10$}

\begin{table}[H]
\scriptsize
\begin{tabular}{lllllllllll}
  \hline
\textbf{Scenario 2} & $\mu_1$ & $\mu_2$ & $\mu_3$ & $\mu_4$ & $\mu_5$ & $\mu_6$ & $\mu_7$ & $\mu_8$ & $\mu_9$ & $\mu_{10}$ \\ 
  \hline
MSE & 0.0033 & 0.0031 & 0.0050 & 0.0097 & 0.0560 & 0.0031 & 0.0036 & 0.0068 & 0.0121 & 0.0629 \\ 
C.I. $90\%$  & 0.88 & 0.90 & 0.93 & 0.90 & 0.92 & 0.88 & 0.83 & 0.87 & 0.90 & 0.86 \\ 
C.I. $95\%$ & 0.95 & 0.97 & 0.97 & 0.94 & 0.98 & 0.96 & 0.93 & 0.92 & 0.94 & 0.89 \\ 
ESS & 779 & 499 & 616 & 695 & 465 & 781 & 548 & 555 & 659 & 457 \\ 
   \hline
   \hline
\end{tabular}
\begin{tabular}{lllllllllll}
  \hline
 & $\phi_1$ & $\phi_2$ & $\phi_3$ & $\phi_4$ & $\phi_5$ & $\phi_6$ & $\phi_7$ & $\phi_8$ & $\phi_9$ & $\phi_{10}$ \\ 
  \hline
MSE & 0.0462 & 0.0239 & 0.0038 & 0.0009 & 0.0003 & 0.0321 & 0.0340 & 0.0035 & 0.0010 & 0.0003 \\ 
C.I. $90\%$ & 0.98 & 0.95 & 0.83 & 0.90 & 0.81 & 0.98 & 0.96 & 0.90 & 0.92 & 0.82 \\ 
C.I. $95\%$ & 0.98 & 0.98 & 0.89 & 0.96 & 0.92 & 1.00 & 0.99 & 0.92 & 0.97 & 0.92 \\ 
ESS & 480 & 385 & 362 & 402 & 319 & 478 & 412 & 369 & 393 & 325 \\ 
   \hline
\end{tabular}
\begin{tabular}{lllllllllll}
  \hline
 & $\sigma_1$ & $\sigma_2$ & $\sigma_3$ & $\sigma_4$ & $\sigma_5$ & $\sigma_6$ & $\sigma_7$ & $\sigma_8$ & $\sigma_9$ & $\sigma_{10}$ \\ 
  \hline
MSE& 0.0082 & 0.0047 & 0.0028 & 0.0018 & 0.0026 & 0.0068 & 0.0053 & 0.0028 & 0.0018 & 0.0027 \\ 
C.I. $90\%$ & 0.95 & 0.97 & 0.85 & 0.89 & 0.81 & 0.96 & 0.98 & 0.91 & 0.93 & 0.84 \\ 
C.I. $95\%$ & 0.97 & 0.99 & 0.95 & 0.96 & 0.89 & 0.98 & 0.99 & 0.96 & 0.96 & 0.90 \\ 
ESS & 283 & 297 & 298 & 300 & 242 & 288 & 317 & 295 & 295 & 236 \\ 
   \hline
\end{tabular}

\begin{tabular}{lllllllllll}
  \hline
 & $\tau_1$ & $\tau_2$ & $\tau_3$ & $\tau_4$ & $\tau_5$  & $\tau_6$ & $\tau_7$ & $\tau_8$ & $\tau_9$ & $\tau_{10}$ \\ 
  \hline
MSE & 0.0112 & 0.0195 & 0.0305 & 0.0440 & 0.0590 & 0.0112 & 0.0195 & 0.0309 & 0.0435 & 0.0590 \\ 
C.I. $90\%$ & 0.78 & 0.85 & 0.77 & 0.74 & 0.78 & 0.79 & 0.79 & 0.81 & 0.77 & 0.76 \\ 
C.I. $95\%$ & 0.83 & 0.85 & 0.84 & 0.80 & 0.82 & 0.82 & 0.87 & 0.84 & 0.81 & 0.80 \\ 
ESS & 520 & 474 & 279 & 261 & 181 & 498 & 459 & 275 & 253 & 164 \\ 
   \hline
\end{tabular}

\begin{tabular}{lllllllllll}
  \hline
 & $s_{300,1}$ & $s_{300,2}$ & $s_{300,3}$ & $s_{300,4}$ & $s_{300,5}$  &$s_{300,6}$& $s_{300,7}$ &$s_{300,8}$ &$s_{300,9}$ & $s_{300,10}$ \\ 
  \hline
MSE & 0.0815 & 0.0915 & 0.1839 & 0.1536 & 0.2546 & 0.0877 & 0.0941 & 0.1639 & 0.1711 & 0.2437 \\ 
C.I. $90\%$ & 0.86 & 0.84 & 0.92 & 0.91 & 0.87 & 0.85 & 0.83 & 0.92 & 0.91 & 0.90 \\ 
C.I. $95\%$ & 0.91 & 0.93 & 0.94 & 0.95 & 0.95 & 0.94 & 0.92 & 0.96 & 0.96 & 0.94 \\ 
ESS & 1073 & 1074 & 1043 & 1007 & 620 & 1087 & 1100 & 1010 & 999 & 628 \\  
   \hline
\end{tabular}
\begin{tabular}{llllll}
  \hline
&$v_{100}$&$v_{200}$&$v_{500}$&$v_{800}$&$v_{900}$ \\ 
  \hline
MSE & 0.0269 & 0.0263 & 0.0190 & 0.0157 & 0.0146  \\ 
C.I. $90\%$ & 0.82 & 0.91 & 0.82 & 0.90 & 0.85 \\ 
C.I. $95\%$ & 0.88 & 0.93 & 0.85 & 0.95 & 0.90 \\ 
ESS & 393 & 406 & 395 & 405 & 406  \\ 
   \hline
\end{tabular}
\caption{MSE estimated using the posterior mode, observed coverage probability of the credible intervals (C.I.) and effective samples size calculated from 2000 posterior draws for selected parameters (Scenario 2).}
\label{tab:sumsim2}
\end{table}

\begin{table}[ht]
\centering
\begin{tabular}{rrrrrrrrrrr}
  \hline
 & $m_1$ & $m_2$ & $m_3$ & $m_4$ & $m_5$& $m_6$ & $m_7$ & $m_8$ & $m_9$ & $m_{10}$ \\ 
  \hline
& 94$\%$& 89$\%$ & 92$\%$ & 93$\%$ & 78$\%$ & 91$\%$ & 88$\%$ & 91$\%$ & 91$\%$ & 81$\%$ \\   
   \hline
\end{tabular}
\caption{Proportion of how often the correct copula family was selected. The selected copula family is the posterior mode estimate of $m_j$ for $j=1, \ldots, 10$ (Scenario 2).}
\label{tab:famsel2}
\end{table}

\end{document}